\documentclass[sigconf]{acmart}
\AtBeginDocument{%
  }

\copyrightyear{2026}
\acmYear{2026}
\setcopyright{cc}
\setcctype{by}
\acmConference[WWW '26] {Proceedings of the ACM Web Conference 2026}{April 13--17, 2026}{Dubai, United Arab Emirates.}
\acmBooktitle{Proceedings of the ACM Web Conference 2026 (WWW '26), April 13--17, 2026, Dubai, United Arab Emirates}
\acmISBN{979-8-4007-2307-0/2026/04}
\acmDOI{10.1145/3774904.3792430}
\acmSubmissionID{rfp2187}

\newcommand{\modelname}{DA-RAG}

\usepackage{algorithm}
\usepackage[noend]{algpseudocode}
\usepackage{pifont}
\usepackage{graphicx}       % for including graphics
\usepackage{svg}
\usepackage[noend]{algpseudocode}
\usepackage{amsmath}
\usepackage{enumitem}

\usepackage{tabularx}
\usepackage{multirow}
\usepackage[table,xcdraw]{xcolor}

\usepackage{booktabs}   % 导入booktabs宏包以获得更美观的表格线
\usepackage{listings}
\usepackage[most]{tcolorbox}
\tcbuselibrary{listings, breakable, skins} 

\definecolor{promptbg}{RGB}{252,250,242}         % 淡米色背景（低干扰）
\definecolor{promptframe}{RGB}{150,120,80}       % 深褐色边框（更有质感）
\definecolor{prompttitlebg}{RGB}{245,230,180}    % 浅米金标题背景
\definecolor{prompttext}{RGB}{30,30,30}          % 深灰文本（柔和但清晰）

\lstdefinestyle{mystyle}{
    language={}, 
    basicstyle=\ttfamily\small, % 非等宽 sans 字体 
    breaklines=true,
    breakatwhitespace=true,
    keepspaces=true,
    showstringspaces=false,
    tabsize=2,
    numbers=none,
    frame=none,
    backgroundcolor=\color{promptbg},
    keywordstyle={},
    commentstyle={},
    stringstyle={},
}
\newtcolorbox{promptbox}[1][]{
    colback=promptbg,
    colframe=promptframe,
    coltitle=black,
    colbacktitle=prompttitlebg,
    fonttitle=\bfseries\ttfamily,
    title style={topsep=1mm, bottomsep=1mm, left=2mm },
    boxrule=0.5pt,
    arc=0mm,
    listing only,
    verbatim=false,
    listing options={style=mystyle},
    #1 % Pass any other tcolorbox options
}

\newcommand{\add}[1]{{\color{black}{#1}}}

\begin{document}

\title{\modelname{}: Dynamic Attributed Community Search for Retrieval-Augmented Generation}


\author{Xingyuan Zeng}
\email{zengxy96@mail2.sysu.edu.cn}
\affiliation{%
  % \institution{Key Laboratory of Intelligent Assessment Technology for Sustainable Tourism}
  \institution{The Technology Innovation Center for Collaborative Applications of 
Natural Resources Data in GBA, MNR}
  \institution{Sun Yat-sen University}
  \city{Zhuhai}
  \country{China}
}

\author{Zuohan Wu}
\email{zh.wu@connect.hkust-gz.edu.cn}
\affiliation{%
  \institution{The Hong Kong University of Science and Technology (Guangzhou)}
  \city{Guangzhou}
  \country{China}
}

\author{Yue Wang}
\email{yuewang@sics.ac.cn}
\affiliation{%
  \institution{Shenzhen Institute of Computing Sciences}
  \city{Shenzhen}
  \country{China}
}

\author{Chen Zhang}
\email{jason-c.zhang@polyu.edu.hk}
\affiliation{%
  \institution{The Hong Kong Polytechnic University}
  \city{Hong Kong}
  \country{China}
}

\author{Quanming Yao}
\email{qyaoaa@tsinghua.edu.cn}
\affiliation{%
  \institution{Tsinghua University}
  \institution{State Key laboratory of Space Network and Communications}
  \institution{Beijing National Research Center for Information Science and Technology}
  \city{Beijing}
  \country{China}
}

\author{Libin Zheng}
\authornote{Corresponding author.}
\email{zhenglb6@mail.sysu.edu.cn}
\author{Jian Yin}
\email{issjyin@mail.sysu.edu.cn}
\affiliation{%
  \institution{Sun Yat-sen University}
  \city{Zhuhai}
  \country{China}
}

\renewcommand{\shortauthors}{Xingyuan Zeng et al.}


\begin{abstract}
% Graph-based Retrieval-Augmented Generation (G-RAG) enhances large language models by grounding them with external knowledge graphs. Nonetheless, existing G-RAG methodologies underutilize graph topology, centering on low-order structural or pre-computed static communities. This limits their performance for dynamic and complex queries.  Thus, we propose DA-RAG, which leverages attributed community search (ACS) to dynamically mine relevant subgraphs for the queried question. DA-RAG captures high-order graph structures, thus retrieving knowledge that is self-complementary. DA-RAG is also equipped with a chunk-layer oriented graph index, which enables efficient multi-granularity retrieval while substantially reducing both computational and economic costs. We evaluate \modelname{} on multiple datasets, demonstrating that it outperforms existing RAG methods by up to 40\% in head-to-head comparisons across four metrics while reducing index construction time and token overhead by up to 37\% and 41\%, respectively.

Owing to their unprecedented comprehension capabilities, large language models (LLMs) have become indispensable components of modern web search engines. From a technical perspective, this integration represents retrieval-augmented generation (RAG), which enhances LLMs by grounding them in external knowledge base. A prevalent technical approach in this context is graph-based RAG (G-RAG). However, current G-RAG methodologies frequently underutilize graph topology, predominantly focusing on low-order structures or pre-computed static communities. This limitation affects their effectiveness in addressing dynamic and complex queries. Thus, we propose DA-RAG, which leverages attributed community search (ACS) to dynamically extract relevant subgraphs based on the queried question. DA-RAG captures high-order graph structures, allowing for the retrieval of self-complementary knowledge. Furthermore, DA-RAG is equipped with a chunk-layer oriented graph index, which facilitates efficient multi-granularity retrieval while significantly reducing both computational and economic costs. We evaluate DA-RAG on multiple datasets, demonstrating that it outperforms existing RAG methods by up to 40\% in head-to-head comparisons across four metrics while reducing index construction time and token overhead by up to 37\% and 41\%, respectively.

% a novel G-RAG architecture, termed \modelname{}. Our approach introduces two primary innovations: 1) it leverages Attributed Community Search (ACS) to dynamically mine relevant subgraphs (communities) online, guided directly by user query semantics; and 2) it 
\end{abstract}



% \received{7 October 2025}
% \received[revised]{7 October 2025}
% \received[accepted]{13 January 2026}

%% information and builds the first part of the formatted document
\setlength{\textfloatsep}{5pt}
\setlength{\floatsep}{5pt}
\setlength{\intextsep}{5pt}
\setlength{\abovecaptionskip}{5pt}
\setlength{\belowcaptionskip}{5pt}

% 如果是双栏文档，还可以添加
\setlength{\dbltextfloatsep}{5pt}
\setlength{\dblfloatsep}{5pt}
\maketitle
\newcommand\webconfavailabilityurl{https://doi.org/10.5281/zenodo.18296495}
\ifdefempty{\webconfavailabilityurl}{}{
\begingroup\small\noindent\raggedright\textbf{Resource Availability:}\\
% please change the following context to include multiple artifacts if necessary, including data, models, code, etc.
The source code of this paper has been made publicly available at \url{\webconfavailabilityurl}.
\endgroup
}
\section{Introduction}
\begin{figure*}[t]
    \centering
    \includegraphics[width=\textwidth]{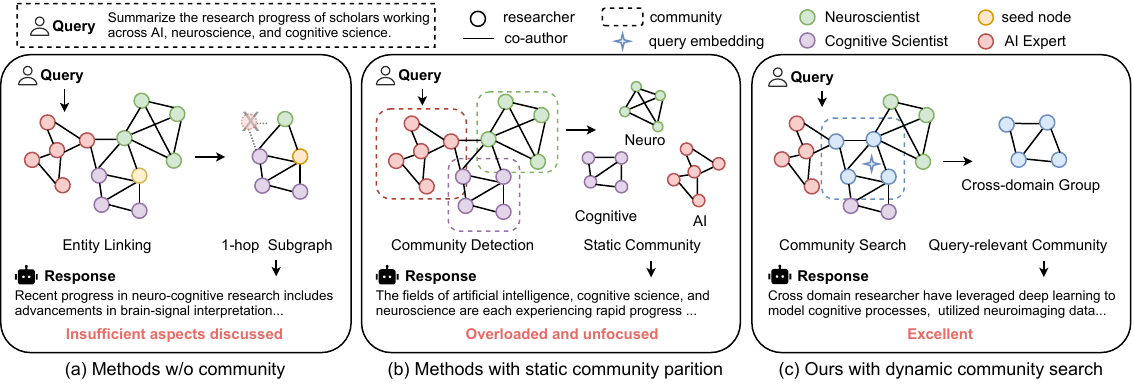}
    \vspace{-12pt}
    \caption{Differences between existing methods and our method. (a) Methods w/o community concern are limited to low-order graph topology, capturing only partial aspects. (b)  Methods with static community partition could return a diverging and unfocused response. (c) Our 
    % dynamic-community-search-based 
    method retrieves a query-relevant subgraph tailored to the question’s need.
    % , generating rich content that discovers its inherent correlations.
    }
    \label{fig:intro}
    \Description{}
    \vspace{-5pt}
\end{figure*}

Retrieval-Augmented Generation (RAG) \cite{rag2020nuerips,rag-survey} has emerged as a prominent technique for enhancing large language models (LLMs). \add{An exemplary RAG application is Microsoft Copilot\footnote{https://copilot.microsoft.com/}, which once sparked a trend of integrating LLMs into web search. On the one hand, for search engines, LLMs can summarize the desired content for users as an indispensable assistant in the modern industry~\cite{DBLP:journals/tsc/XiongBLLDWYH24}. On the other hand,} by incorporating relevant context retrieved from external knowledge sources, RAG enables LLMs to generate accurate, timely, and domain-specific responses without altering their underlying parameters. Among the various emerging RAG paradigms, graph-based RAG (G-RAG) \cite{GraphRAG,hippoRAG, Grag-survey} is gaining popularity, due to its advantage of employing a graph to represent the relationships among data entities. Compared to traditional RAG approaches \cite{rag2020nuerips,rag-survey,rag-sur2}, which treat documents or text chunks as discrete and independent units, G-RAG  captures the complex semantic relationships within the data. This helps to retrieve a knowledge collection that is inherently correlative and complementary.

% due to its unique advantages in modeling and utilizing interconnected information among data\cite{Grag-survey}. Unlike traditional RAG approaches\cite{rag2020nuerips,rag-survey,rag-sur2}, which treat documents or text chunks as discrete and independent units, G-RAG employs an explicit graph structure to represent the inherent relationships among data entities\cite{Grag-survey}. This approach captures the complex relationships and hierarchical structures within the data via retrieving subgraphs from the local knowledge base (as a graph)\cite{subgraphsfromkg}. 

Regarding the local knowledge base as a graph, G-RAG approaches generally retrieve a high-quality knowledge subset by fetching a subgraph that is both relevant to the query (question) and internally densely connected. With subgraph retrieval as the core technical module \cite{Grag-survey}, current G-RAG methodologies still encounter the significant challenge of  \textbf{coarse-grained exploration of graph topology}. 
Despite the potent representational power and rich semantic topology inherent in graph-structured data \cite{Graphrepresentationpower}, the current utilization of graph structures in G-RAG methods remains straightforward.
% Graph-structured data possesses potent representational power and rich semantic topology \cite{Graphrepresentationpower}, offering substantial opportunities for deep exploration. However, the current utilization of graph structures in G-RAG methods remains straightforward.
For instance, Approaches such as ToG \cite{ToG}, MindMap \cite{MindMap}, DALK \cite{DALK}, and LightRAG \cite{lightRaG} predominantly leverage immediate adjacency relationships to retrieve direct neighbors, local pathways, or n-hop subgraphs. HippoRAG \cite{hippoRAG} and GNN-RAG \cite{GNN-RAG} employ graph algorithms like PageRank \cite{page1999pagerank} or Graph Neural Networks \cite{GNN} to assess node importance or identify relevant paths. These approaches focus on low-order structural information confined to pairwise or path-level connectivity, failing to capture the higher-order structural information \cite{In-depthGrag} inherent in graph data. 

% These methodologies rely on low-order structural information confined to pairwise connectivity, failing to capture the higher-order structural information\cite{In-depthGrag} inherent in graph data. To better exploit graph topology, 

% HippoRAG\cite{hippoRAG} and GNN-RAG\cite{GNN-RAG} employ graph algorithms like PageRank\cite{page1999pagerank} or Graph Neural Networks (GNNs)\cite{GNN} to assess node importance or identify relevant paths. While these methods represent an advancement, they remain centered on node or path-level properties. 
% Consequently, they may overlook crucial collective patterns within graph data, particularly communities, which can encapsulate abstract concepts or latent topics/themes that gather disparate information fragments. 

Microsoft's GraphRAG \cite{GraphRAG} and ArchRAG \cite{ArchRAG} represent initial efforts to explore deeper graph topology by treating subgraphs as communities \cite{community,fortunato2010community}.  They apply graph clustering techniques \cite{leiden} to partition large-scale knowledge graphs into distinct communities \textit{a priori} (in the offline index construction process), and query over such 
fixed (static) clusters online.  Nevertheless, such static and pre-computed communities may not fit the diverse queries that users issue to LLM. 
For example, consider a manager asking to summarise the employee cooperation among the departments for a certain project. This requires information cutting across predefined community boundaries, when the employee entities are offline clustered according to the department membership. Therefore, \textbf{how to \underline{dynamically mine high-order} subgraph/community patterns subject to the diverse user queries}, remains a pivotal unresolved challenge.

% \textbf{II) Computational Overhead and Economic Costs from Complex Graph Operations:} In order to offer both summarizing and domain-specific knowledge, existing G-RAG approaches generally build a full graph hierarchy \cite{GraphRAG, ArchRAG,lightRaG} with inter-links among different graph layers. This icurs either frequent graph operations (e.x., full-graph clustering\cite{Louvain,leiden}) or invocation of Large Language Models (LLMs) \cite{GraphRAG,ArchRAG,ToG,Cot}. These computational and Economic challenges pose significant barriers to the widespread adoption and practical deployment of G-RAG in large-scale, real-world applications\cite{Grag-survey}.

% enable multi-granularity sub-graph retrieval 
% \cite{GraphRAG,ArchRAG

% Notably, operations such as full-graph clustering\cite{Louvain,leiden} for community detection\cite{CD} can significantly escalate computational demands. Furthermore, the frequent invocation of Large Language Models (LLMs) by some G-RAG methods\cite{GraphRAG,ArchRAG,ToG,Cot}  introduce substantial economic costs. 

% In response to these challenges, we propose \modelname{} (\underline{D}ynamic \underline{A}ttributed community search for RAG) whose key insight is to \underline{adaptively} mine the entity community from the local KG, according to the query's semantics, as demonstrated in Figure \ref{fig:enter-label}. By applying the paradigm of embedding-attributed community search from the area of graph analytics to RAG's subgraph retrieval, it retrieves knowledge that is self-complementary, inherently rich, and related to the question.

In response to these challenges, we propose \modelname{} (\underline{D}ynamic \underline{A}ttributed community search for RAG).
To the best of our knowledge, our work is the first to introduce and adapt the concept of Attributed Community Search (ACS) from graph analytics to serve the specific needs of RAG.
Specifically, we reframe the subgraph retrieval task in G-RAG as an embedding-attributed community search problem. This paradigm shift, illustrated in Figure \ref{fig:intro}, enables \modelname{} to dynamically identify a community from the knowledge graph that is both structurally cohesive and semantically guided by the query. 
Furthermore, to realize cost-effective online retrieval, DA-RAG is equipped with a chunk-layer oriented graph index, which primarily mirrors the logical structure of source documents by treating text chunks as graph nodes. The index further grows another two graph layers, considering the similarity and inherent connections among entities, respectively. In this way, the subgraph retrieval flows from the chunk layers to the two grown fine-grained layers. 
To summarize, our key contributions are as follows:
\begin{enumerate}
 \item We pioneer a new subgraph retrieval paradigm for RAG by formulating an Embedding-Attributed Community Search (EACS) problem, which adapts ACS to dynamically retrieve structurally cohesive and semantically relevant subgraphs.

 \item We design an efficient, multi-granularity Chunk-layer Oriented Graph Index that eliminates expensive clustering, reducing indexing costs while supporting queries at various levels of detail.

 \item We demonstrate through extensive experiments that DA-RAG significantly outperforms state-of-the-art baselines in both response quality and end-to-end efficiency (indexing and retrieval).
\end{enumerate}

\begin{figure*}[!t]
    \centering
    \includegraphics[width=\textwidth]{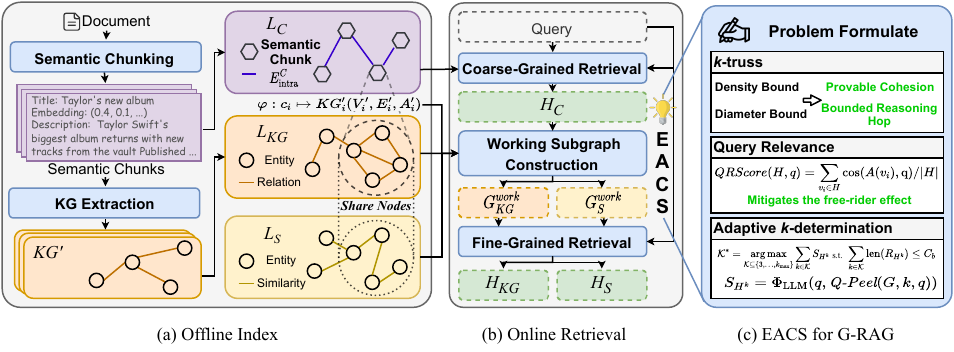}
    \vspace{-15pt}
\caption{
Overview of the DA-RAG framework: (a) Offline Indexing creates a novel graph index from source documents, comprising a high-level layer ($L_C$) and two granular layers ($L_{KG}$ and $L_S$). (b) Online Retrieval employs a coarse-to-fine strategy. (c) EACS Formulation defines the subgraph retrieval in G-RAG as the Embedding-Attributed Community Search (EACS), ensuring \textbf{provable cohesion, bounded reasoning hops}, and \textbf{mitigates free-rider effects.}}
    \label{fig:overview}
    \Description{}
    \vspace{-5pt}
\end{figure*}

% \section{\wzh{DA-RAG (with its full name)}}
% \wzh{zuohan: here we organize the methodology into one section, or two sections for each stage.}
\vspace{-9pt}
\section{Overview of DA-RAG}
% \wzh{Here, introduce the pipeline of DA-RAG with a \textbf{figure}. Also, the input/output, e.g., 
% ``Given the document corpus, DA-RAG will first construct a dual-level graph index, considering both xxx and xxx. (Discuss the output of the offline stage) When a query, embedded in $q$, arrives, we (briefly introduce the process given a query, especially formalize the input/output) ''}
% \wzh{\textit{should rewrite, here just depicts the structure}. }

As illustrated in Figure~\ref{fig:overview}, DA-RAG operates in two stages: a one-time \textbf{Offline Indexing} phase (Figure~\ref{fig:overview}(a)) and a dynamic \textbf{Online Retrieval} process (Figure~\ref{fig:overview}(b)). At the heart of its online retrieval lies a new subgraph search paradigm, which we formulate as the \textbf{Embedding-Attributed Community Search (EACS)} problem (Figure~\ref{fig:overview}(c)). This integrated framework enables the retrieval of contextually rich and structurally coherent subgraphs, effectively addressing key challenges in G-RAG.
 
In the offline phase, DA-RAG processes an input document corpus to engineer a graph structure, termed the Chunk-layer Oriented Graph Index. Its cornerstone, the Semantic Chunk Layer ($L_C$), comprises nodes of semantic text chunks that provide high-level context by preserving the document's narrative structure, thereby \textbf{avoiding expensive graph clustering }used in other methods \cite{GraphRAG,ArchRAG}. This layer is hierarchically linked to two fine-grained perspectives, the Knowledge Graph ($L_{KG}$) and Similarity ($L_S$) layers to form a dual-level, multi-perspective index, as detailed in Section~\ref{sec:offline index}.

During the online retrieval phase for queries, DA-RAG employs a \textbf{coarse-to-fine strategy}. First, a coarse search at the Chunk layer ($L_C$) identifies an initial community, $H_C$, which provides abstract, high-level context for the query \cite{In-depthGrag}. This community then guides the pruning of the Knowledge Graph Layer ($L_{KG}$) and Similarity Layer ($L_S$), allowing for a final, fine-grained search within the resulting subgraphs. As a result, the process retrieves two detailed communities, \(H_{KG}\) and \(H_S\), each offering different perspectives. 
For more details, please refer to Section~\ref{sec:Retrieval Workflow}.

Particularly, central to our online retrieval process is Embedding Attributed Community Search (EACS), detailed in~Section~\ref{sec:EACS}, 
a novel query-guided subgraph retrieval paradigm,
which guarantees \textbf{provable cohesion} and a \textbf{bounded reasoning hop} via k-truss, \textbf{mitigates the ``free-rider effect''} \cite{freerider} through a custom relevance score, and adaptive determination of $k$ for $k$-truss.

% \subsection{\wzh{Chunk-layer Oriented Graph Index (offline index construction)}}

\vspace{-10pt}
\section{Offline Index}
\label{sec:offline index}

The first stage of the RAG standard workflow involves organizing the knowledge base \cite{rag-survey}. 
% To create a comprehensive multi-granularity index while minimizing computational overhead, 
Our approach deeply leverages the inherent structure of standard G-RAG workflows \cite{In-depthGrag}. We choose semantic chunking \cite{semchunk} over fixed-length methods \cite{GraphRAG}. This choice ensures that each chunk effectively captures a coherent segment of the document's logic and narrative, yielding a set of semantic chunks denoted as $\{c_i\}_{i=1}^N$.
We then conduct knowledge graph extraction \cite{GraphRAG} for each semantic chunk $c_i$ to construct a local knowledge graph, denoted by the mapping $\varphi: c_i \mapsto KG_i'(V_i', E_i', A_i')$. 

Our core insight is that the combination of semantic chunking and knowledge graph extraction gives rise to an \textbf{emergent semantic hierarchy}. Each semantic chunk $c_i$ serves as a high-level abstraction over the detailed entities and relations in its corresponding graph $KG_i'$. This approach provides a cost-effective way to form a hierarchical structure, avoiding the need for computationally expensive methods like graph clustering.

To further enrich the index and mitigate the common issue of graph sparsity \cite{sparsity}, we incorporate semantic similarity edges as proposed in previous work \cite{DALK}. These edges are maintained in a separate Similarity Layer ($L_S$) to preserve the unique topology and relational semantics \cite{RGCN} of the Knowledge Graph Layer ($L_{KG}$) . Summarizing the above insights, we propose a three-layer synergistic index as detailed in the following:

% To further enrich the index and mitigate the common issue of graph sparsity \cite{sparsity}, we follow the recent work \cite{hippoRAG} to incorporate semantic similarity edges. 
% However, a key architectural decision (\wzh{what decision? You mean the novelty of our design?}) is to isolate these edges into a dedicated Similarity Layer ($L_S$) rather than merging them into the Knowledge Graph Layer ($L_{KG}$). 
% This separation is vital to preserve the distinct topology and explicit relational semantics of the KG, which can be obscured when mixing heterogeneous edge types \cite{RGCN}. 

% We then additionally build another graph layer with the chunks as nodes, in pursuit of multi-granularity information access (over the chunk layer and the original KG layer) for the text corpus. The two layers are summarized below.

% Knowledge graphs extracted solely from text corpora often suffer from sparsity  \cite{sparsity}, meaning the explicit relations ($E$) may not adequately support complex knowledge reasoning.

\textbf{Semantic Chunk Layer ($L_C$):} At a coarse-level, each node $v_i^c$ in this layer represents a semantic text chunk $c_i$. We use an LLM to generate a title and a concise description for each chunk, %serving as node attributes, 
which is further turned into a vector representation. Crucially, this embedding process utilizes the same embedding model employed to generate the query embedding, ensuring consistency in the vector space. Inspired by hierarchical clustering approaches~\cite{leiden}, if the knowledge subgraphs extracted from two distinct text chunks, $c_i$ and $c_j$ ($i \neq j$), are connected by a relation in the global KG (i.e., there exists a relation $(u,w) \in E$ such that one entity is in $V_i'$ and the other is in $V_j'$), we add an edge $E_{\text{intra}}^C$ between their corresponding chunk nodes $v_i^c$ and $v_j^c$ in this layer as 

% \begin{equation*}
% % 使用 equation* 环境，不带编号
% E_{\text{intra}}^C = 
%   \left\{ 
%     (v_i^c, v_j^c) \mid
%     \begin{gathered} % 创建一个顶部对齐的子环境
%       \exists (u,w) \in E \text{ s.t. } \\
%       ((u \in V_i' \land w \in V_j') \\ 
%       \lor (u \in V_j' \land w \in V_i'))  \\ 
%       \text{and }  i \neq j                                  
%     \end{gathered}
%   \right\}.
% \end{equation*}

% \begin{equation*}
% E_{\text{intra}}^C = \left\{ (v_i^c, v_j^c) \mid (i,j) \in P \right\},
% \end{equation*}
% \begin{equation*}
% P = \left\{ (i,j) \mid i \neq j \text{ and } \exists u \in V_i', w \in V_j' \text{ s.t. } (u,w) \in E \right\}.
% \end{equation*}

\vspace{-10pt}
\begin{equation*}
\begin{gathered}
E_{\text{intra}}^C = \left\{ (v_i^c, v_j^c) \mid (i,j) \in P \right\}, \\
P = \left\{ (i,j) \mid i \neq j \text{ and } \exists u \in V_i', w \in V_j' \text{ s.t. } (u,w) \in E \right\}.
\end{gathered}
\end{equation*}
% \vspace{-5pt}

\textbf{Knowledge Graph Layer ($L_{KG}$):} As a fine-grained layer, $L_{KG}$ refers the global knowledge graph extracted from the corpus, primarily containing entity nodes ($V$), relations between them ($E$), and associated entity embeddings ($A$). 

The above two layers are connected through inter-layer links based on the mapping $\varphi$. Specifically, each chunk node $v_i^c \in L_C$ is linked to all entity nodes $u \in V_i'$ within its corresponding knowledge subgraph.
\[E_{\text{inter}} = \left\{ (v_i^c, u) \mid v_i^c \in L_C, u \in V_i' \right\}.\]

\textbf{Similarity Layer ($L_S$):} As another fine-grained layer complementary to $L_{KG}$, this layer comprises the same nodes (entities) as $L_{KG}$ but with edges defined by semantic proximity. 
% Specifically, for node pairs whose embedding similarity is larger than a threshold \libin{(whose sensitivity analysis is provided in Appendix C, and is accordingly set to 5)}, we add an edge representing semantic similarity in a dedicated Similarity Layer ($L_S$). 
We employ the $k$-Nearest Neighbors (KNN) algorithm to build the Similarity Layer ($L_S$). In this layer, we connect each node to its top-$k$ similar neighbors based on embedding similarity. A sensitivity analysis to identify the optimal value of $k_{neighbor}$ is provided in Section~\ref{sec:Further Analysis}.
% and is accordingly set to 5. 
Additionally, we create interlinks between this layer and the chunk layer, similar to our previous approach.

Overall, our offline process builds a dual-level (Semantic Chunk Layer $L_C$ + Knowledge Graph Layer $L_{KG}$) and multi-perspective (structural relations in $L_{KG}$ + semantic similarity in $L_S$) index.  
% This structure employs a two-layer architecture for efficient coarse-to-fine mapping and mitigates the KG sparsity problem by incorporating semantic similarity links between entities. 

% This comprehensive index not only lays the foundation for efficient retrieval but also inherently reduces the complexity of global processing due to its hierarchical nature, helping to control computational costs. Next, we describe how this index facilitates efficient, multi-granularity subgraph retrieval during the online phase.

\vspace{-5pt}
\section{Online Retrieval Workflow}
\label{sec:Retrieval Workflow}
% \wzh{add this section to the online part of the pipeline from a coarse perspective, like processing the query, retrieval (cross-grained) and get a subgraph, refine the subgraph, output the answer.}

% In this section, we describe the index framework phase by phase as follows.
Given the indexed corpus, we develop an efficient, coarse-to-fine retrieval strategy that narrows the search range when the queries arrive. Specifically, this strategy progressively reduces the search space by breaking down the overall retrieval task into a series of sub-retrieval steps. Each step is specified as a subgraph retrieval problem, aiming to identify the optimal subgraph from different graph layers. In this section, we will first outline the overall workflow of our proposed strategy. %Particularly, we would delve into the detailed formulation of the subgraph retrieval problem in the next Section~\ref{sec:EACS} after a holistic introduction in this section.

% The specifics of each phase are detailed below.

% The knowledge retrieval of DA-RAG proceeds as shown in Figure~\ref{fig:overview}(b).
\textbf{Coarse-Grained Retrieval.}
Given the embedding $q$ derived from the user's natural language query, we initiate the retrieval process by performing EACS (a subgraph retriever to be detailed in Section~\ref{sec:EACS}) on the coarse level.
We specifically operate at the Semantic Chunk Layer \( L_C \) since this approach is less computationally intensive.
% , i.e., the Semantic Chunk Layer $L_C$, since it is less computationally demanding. 
It identifies and generates an attribute community $H_C \subseteq L_C$, $H_C$ providing a contextual anchor for subsequent fine-grained exploration.

\textbf{Working Subgraph Construction.}
Leveraging the inter-layer connections $E_{\text{inter}}$, the retrieved chunk community $H_C$ guides the identification of relevant entities within the Knowledge Graph Layer $L_{KG}$ and Similarity Layer $L_S$. Specifically, we collect all entity nodes that are connected to any chunk node within $H_C$, forming the entity set 
$V_{\text{work}} = \{ u \in V \mid \exists v_i^c \in H_C \text{ s.t. } (v_i^c, u) \in E_{\text{inter}} \}.$
Based on this entity set, we induce two working subgraphs: $G_{KG}^{\text{work}}$ for entities $V_{\text{work}}$  within the layer $L_{KG}$, and $G_{S}^{\text{work}}$  for them within the layer $L_S$.
% Mathematically: G_KG^work = (V_work, E[V_work]), G_S^work = (V_work, E^S[V_work]) where E[V_work] and E^S[V_work] are induced edge sets.

\textbf{Fine-Grained Retrieval.}
For a further refinement before generation, we execute EACS again on these significantly smaller working subgraphs $G_{KG}^{\text{work}}$ and $G_{S}^{\text{work}}$. This step aims to identify fine-grained communities: $H_{KG}$ within $G_{KG}^{\text{work}}$ (representing relevant entities connected by explicit relations) and $H_S$ within $G_{S}^{\text{work}}$ (representing relevant entities connected by semantic similarity). These communities constitute fine-grained knowledge units highly relevant to the query and internally cohesive.

\begin{figure}[t!]
\centering
\includegraphics[width=0.46\textwidth]{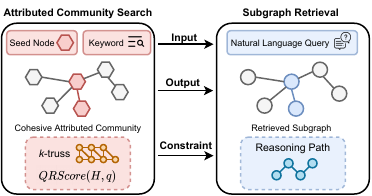}
\vspace{-5pt}
\caption{An illustration mapping the G-RAG subgraph retrieval task to the Attributed Community Search problem. 
% The analogy is established through three components: input, output, and constraint.
}
\label{fig: analogy table}
   \Description{}
   \vspace{-6pt}
\end{figure}
\vspace{-5pt}
\section{EACS: Key Module for Online Retrieval}
\label{sec:EACS}

% \subsection{Problem Mapping}
% Now, the leaving part is detailing Equation xxx or process xxx
The only thing we leave in the online stage is to decide the implementation of subgraph retrieval. Generally, subgraph retrieval of G-RAG can be represented as $G = \text{G-Retriever}($q$, \mathcal{G})$, where $G$ is a subgraph of the local database $\mathcal{G}$ and $q$ is the user query. For this process, we observe that there exists a rational mapping from the area of  G-RAG to attributed community search \cite{ACS,ATC}, as shown in Figure \ref{fig: analogy table}.
% ~\wzh{Specifically, (more details about figure 3, I cannot understand it.)}
Specifically, the natural language query ($q$) provides semantic guidance, akin to the keywords and seed nodes in ACS. The retrieved subgraph, our desired output, corresponds to the cohesive attributed community. Crucially, the implicit requirement for a ``reasoning path'' within the subgraph imposes a structural constraint that mirrors the cohesiveness metrics (e.g., $k$-truss) used in ACS to ensure the community is tightly connected ~\cite{ACS-survey}. 
We then novelly formulate the embedding-attributed community search problem as below, which is expected to capture the high-order semantic connections among nodes in the knowledge graph $\mathcal{G}$. 

% \wzh{why not use a subsetion ``EACS definition''}
% \noindent \textbf{Problem: Embedding-Attributed Community Search (EACS) for G-RAG}: \\
\vspace{-5pt}
\subsection{EACS Definition}
Given a knowledge graph $G(V,E,A)$, natural language query $q$, a $k$-truss \cite{trusses} parameter $k$ (value will be determined at the end of this section), the problem of attributed community search returns a subgraph $H \subseteq G$ satisfying the following properties:

\begin{enumerate}
\item \textbf{Structure cohesiveness}: $H$ is a connected $k$-truss;
\item \textbf{Query Relevance}: The query relevance score $QRScore(H, q)$ (Equation~\eqref{eqScore}) is maximized.
\item \textbf{Maximality}: There does not exist another subgraph $H'~\supset H$ satisfying the above properties.
\end{enumerate}

% We analyze the complexity of Q-Peel in Appendix A.4.~\wzh{this sentence seems redundant, at least, Q-Peel is what?}
% We now dive into the details of why the community search problem goes as above to suit G-RAG.
% \wzh{Three mentioned properties constrained in EACS ensure the following distinct benefits for G-RAG: Provable Cohesion,Bounded Reasoning Hop and Mitigates the free-rider effect}

By design, these constraints provide the G-RAG pipeline with three critical advantages: \textbf{1) Provable Cohesion}, ensuring the retrieved context is thematically consistent; \textbf{2) Bounded Reasoning Hops}, which limits the reasoning space and prevents semantic drift; and \textbf{3) Mitigation of the Free-Rider Effect}, filtering out noisy or irrelevant nodes. In the subsequent analysis, we will deconstruct each of these constraints, demonstrating precisely the mechanism by which each confers its claimed advantage.
% We now dive into the details of why the community search problem goes as above to suit G-RAG.

% \textbf{Input Formulation:} First of all, we treat the local knowledge base as an undirected and attributed graph $G(V,E, A)$ where $V$ is the set of nodes (entities), $E$ is the edges (relations) among nodes, and furthermore each node $v\in V$ is associated with some attributes as $A(v)$.  In contrast to traditional community search works which treat attributes as discrete keyword sets \cite{StructureMeetsKeywords,KCcS}, we represent the node attributes $A(v)$ as embeddings to support the fast retrieval of RAG.  Likewise, the query $q$ itself is also represented by an embedding vector $\mathbf{q}$ to capture its semantics. This embedding-centric formalization naturally aligns with the RAG paradigm \cite{Grag-survey}, facilitating seamless integration with various G-RAG systems. \wzh{This paragraph is hard to follow, should be a part of the pipeline? That's why we should clearly formalize G-RAG before and here we can directly recall the input/output.}

\vspace{2pt}
\textbf{Constraint $H$ is a $k$-truss.} 
The $k$-truss constraint serves as the structural cornerstone for realizing two of our stated advantages: \textbf{Provable Cohesion} and \textbf{Bounded Reasoning Hops}. Its mechanism operates through two fundamental graph-theoretic properties.
First, by definition, a $k$-truss requires every edge to be part of at least $k-2$ triangles. This condition establishes a lower bound on density (we prove in Appendix A.1), which is the direct mechanism for Provable Cohesion, ensuring that the retrieved context is composed of thematically related concepts.
Second, a connected k-truss has a guaranteed upper bound on its diameter, specifically $\lfloor \frac{2|V|-2}{k} \rfloor$ for a subgraph with $|V|$ vertices \cite{trusses} (a concise proof is presented in Appendix A.2). This property directly translates into Bounded Reasoning Hops by imposing a finite upper limit on the path length between any two nodes. This restriction effectively curtails the reasoning space, preventing the semantic drift.
% Beyond its theoretical appeal, the choice of k-truss is pragmatic. It strikes an effective balance between structural integrity and computational tractability \cite{ACS-survey}, offering stronger cohesion than k-cores while avoiding the prohibitive computational cost associated with finding $k$-cliques.
% \textbf{Constraint $H$ is a $k$-truss.} In G-RAG systems, it is desired that  retrieved subgraph is dense \cite{multipath} and the graph diameter \cite{diameter,Curvature} is bounded. This reveals that the nodes in the subgraph are highly correlated to one another and the reasoning path \cite{ToG,Cot} for any pair of nodes is length-bounded. For this objective, we introduce the $k$-truss \cite{trusses} structure to define the target subgraph format. In particular, the $k$-truss requires that each edge in the subgraph be involved in at least $k-2$ triangles, a feature that ensures a tight relationship among the nodes in the subgraph, manifested as a determined lower bound on density (we prove in the Appendix A.1). A connected k-truss subgraph possesses a definite upper bound on its diameter; specifically, for any connected k-truss with $|V|$ vertices, its diameter is at most $\lfloor \frac{2|V|-2}{k} \rfloor$ \cite{trusses} (we present concise alternative proof in Appendix A.2). 

\vspace{2pt}
\textbf{Constraint Query Relevance.} The ``free-rider effect'' in graph analysis refers to the inclusion of nodes in a retrieved community primarily due to their structural connectivity \cite{freerider}, despite lacking direct relevance to the query's core intent. Within G-RAG, such free-rider nodes can introduce noise and dilute the contextual information provided to the LLM, potentially degrading the quality and relevance of its generated responses \cite{lossinmiddle}. Thus, we wish the retrieved community to avoid this effect. For this objective, we define a Query Relevance Score, which allows us to retrieve subgraphs that are closer to the query in the embedding space. Given a subgraph $H \subseteq G$ and an embedding function $f_{\text{embed}}(\cdot)$, the community semantic similarity of $H$, denoted as $QRScore(H,q)$, is the average similarity between nodes in the community and the query $q$:
\begin{equation}\label{eqScore}
    QRScore(H, q) = \frac{\sum_{v_i \in H} \cos(A(v_i), f_{\text{embed}}(q))}{|H|} \,.  
\end{equation}
The EACS formulation, by optimizing for $QRScore$ within a $k$-truss structure, inherently \textbf{mitigates the free-rider effect}, as proved in Appendix A.3.

\vspace{2pt}
\textbf{Adaptive $\mathbf{k}$-determination}
The parameter \( k \) of \( k \)-truss for EACS controls the density and extent of the communities identified during both retrieval steps. 
Recognizing that the optimal  \( k \) is inherently query-dependent, our DA-RAG framework infers the optimal $k$ per query as described below.
%incorporates an adaptive mechanism that dynamically determines the optimal community best aligned with the query's informational needs.

We first generate a set of candidate communities, \( \{H^k\}^{k_{\text{max}}}_{k=3} \), by varying \( k \) over a feasible range. Subsequently, we leverage an LLM to perform a joint evaluation and summarization for each candidate. 
% This mechanism is seamlessly integrated into the community report generation phase of the G-RAG pipeline \cite{GraphRAG}. 
The process for each candidate community $H^k$ is formally represented as
$ S_{H^k}, R_{H^k} = \Phi_{\text{LLM}}(q, H^k) $.
Here, the LLM function $\Phi_{\text{LLM}}$ jointly generates two outputs based on the user query $q$ and candidate community $H^k$: a relevance score $S_{H_k}$ and a community report $R_{H_k}$. 
% The prompt we used here can be found in Appendix B.
% Finally, these communities and reports are ordered by \( S_{H_k} \) and truncated to fit within the context window limit. This strategy ensures that the most relevant information is prioritized and presented to the model for final answer generation.
Finally, an optimal subset of the community is selected to maximize total relevance while adhering to the context budget $C_{b}$. The optimal set of community indices, $\mathcal{K}^*$, is determined by:
\[\mathcal{K}^* = \underset{\mathcal{K} \subseteq \{3, \dots, k_{\text{max}}\}}{\arg\max} \sum_{k \in \mathcal{K}} S_{H^k} \quad \text{s.t.} \quad \sum_{k \in \mathcal{K}} \text{len}(R_{H^k}) \le C_{b}.\]
We solve this via a greedy strategy: communities are ranked by their scores $S_{H^k}$ and packed into the context sequentially until the budget is exhausted. This ensures the most relevant information is prioritized for final answer generation.

\vspace{-5pt}
\subsection{EACS Solution}
% \label{Algorithm}\wzh{also a subsection}
% To address the EACS problem, we propose an efficient algorithm, Q-Peel, based on a multi-stage peeling strategy. The algorithm begins by extracting the maximal $k$-truss subgraph from the given attributed graph (Line 1), by directly invoking the established $k$-truss decomposition algorithm \cite{TrussDecomposition}. For each connected component within this $k$-truss, we refine its nodes by removing those that are not relevant to the query (Lines 2-5). Specifically, nodes are sorted as a queue in the ascending order of their relevance to $q$,  via $\textsc{SortByRelevance}(S,q)$.  We then pop the nodes one by one, and accordingly remove it from the subgraph when two conditions are satisfied (Line 6,  verified using $\textsc{IsValidImprovement}$): (i) the refined subgraph remains a connected $k$-truss, and (ii) the query relevance score ($QRScore$) strictly improves after removal. The refinement continues until the newest popped node could not be removed from the subgraph.
% Finally, after all connected components have been refined (Line 9-10), the algorithm returns the community $H$ that yields the highest $QRScore$ among all candidates processed. We analyze the complexity of Q-Peel and present its pseudocode in Appendix A.5.

\begin{algorithm}[t]
\caption{Q-Peel (Query-aware Peeling)}
\label{al:1}
\small
\begin{algorithmic}[1]
\Require $G$: Undirected attributed graph, $q$: Query embedding, $k$: Target k for k-truss
\Ensure $H$: Optimal community
\State $T_k \gets$ maximal $k$-truss subgraph of $G$
\State$C \gets$ connected components in $T_k$
; $updated \gets$ \textbf{true}
\ForAll{component $c$ in $C$} $S \gets$ nodes in $c$ 
  \While{updated} $updated \gets$ \textbf{false}
    \ForAll{$v$ in \Call{SortByRelevance}{$S, q$}} 
    \State $S' \gets S \setminus \{v\}$
      \If{\Call{IsValidImprovement}{$S', S, q$}}
        \State $S \gets S'$; $updated \gets$ \textbf{true}
        \State \textbf{break}
      \EndIf
    \EndFor
  \EndWhile
  \If{\Call{QRScore}{$H', q$} $<$ \Call{QRScore}{$S, q$}}
  \State $H' \gets S$
 \EndIf
\EndFor
\State \Return $H \gets H'$
\end{algorithmic}
\end{algorithm}

\add{We prove that EACS is an NP-hard problem (see Appendix~\ref{ap:a4} for a detailed proof).} Regarding such hardness, we propose an efficient heuristic named \textbf{Q-Peel}, an efficient multi-stage peeling algorithm to solve the EACS problem. The algorithm operates in three main phases. First, it prunes the input graph by extracting the maximal $k$-truss subgraph using a standard decomposition algorithm \cite{TrussDecomposition}. Second, it refines each connected component of this $k$-truss via an iterative peeling process. Specifically, nodes within a component are sorted in ascending order of their relevance to the query $q$. The algorithm then attempts to sequentially remove nodes, starting from the least relevant. A node is removed if it meets two conditions: it keeps the connected \( k \)-truss structure intact and improves the \( QRScore \) from Equation~\eqref{eqScore}. After processing all components, Q-Peel returns the subgraph with the highest \( QRScore \). The overall Q-Peel is illustrated in Algorithm~\ref{al:1}, which shares \add{a worst-case time complexity of $O(m^{1.5} + cn^2t)$ and a space complexity of $O(n+m)$, where $n$ and $m$ are the number of nodes and edges in the input graph, respectively. Detailed complexity proof can be found in Appendix~\ref{ap:a5}}.

\begin{table*}[ht!]
\centering
\caption{Head-to-head win rates for our proposed DA-RAG versus baseline methods are reported as average percentages (± standard deviation) over five experiments; results $\ge$ 50\% indicate that DA-RAG outperforms the baseline. In the table, GLightRAG, HLightRAG, and MLightRAG refer to three different variants of LightRAG, namely the Global, Hybrid, and Mix versions, respectively. 
Similarly, LGraphRAG and GGraphRAG correspond to the Local and Global variants of GraphRAG.}
\renewcommand{\arraystretch}{1.2}
\setlength{\tabcolsep}{4pt}
\resizebox{\textwidth}{!}{
\begin{tabular}{lcccccccccccc}
\toprule
\multirow{2}{*}{\textbf{Win Rates of Comparison}} & \multicolumn{4}{c}{\textbf{Agriculture}} & \multicolumn{4}{c}{\textbf{Mix}} & \multicolumn{4}{c}{\textbf{News Articles}} \\
\cmidrule(lr){2-5} \cmidrule(lr){6-9} \cmidrule(lr){10-13}
 & \textbf{Comp.} & \textbf{Div.} & \textbf{Emp.} & \textbf{Over.} & \textbf{Comp.} & \textbf{Div.} & \textbf{Emp.} & \textbf{Over.} & \textbf{Comp.} & \textbf{Div.} & \textbf{Emp.} & \textbf{Over.} \\
\midrule
\rowcolor[HTML]{F2F2F2}
\multicolumn{13}{l}{\textbf{Inference-only}} \\
DA-RAG vs Zero-shot & 97.6(±0.4) & 95.7(±0.5) & 95.4(±0.5) & 92.8(±0.7) & 95.9(±0.4) & 92.8(±1.0) & 94.1(±0.4) & 94.7(±0.5) & 96.7(±0.6) & 95.5(±0.3) & 97.9(±0.3) & 95.8(±0.8) \\
DA-RAG vs CoT & 90.9(±1.2) & 94.7(±1.3) & 91.5(±1.1) & 90.8(±1.5) & 89.4(±1.4) & 87.8(±0.9) & 90.3(±1.0) & 90.0(±1.1) & 91.6(±1.0) & 90.2(±0.8) & 91.0(±1.1) & 91.2(±1.3) \\
\midrule
\rowcolor[HTML]{F2F2F2}
\multicolumn{13}{l}{\textbf{Retrieval-only}} \\
DA-RAG vs BM25 & 93.7(±1.1) & 90.9(±0.8) & 93.6(±0.9) & 90.2(±1.0) & 92.8(±0.9) & 91.7(±1.0) & 93.0(±0.8) & 93.3(±0.9) & 92.1(±1.2) & 91.8(±1.0) & 92.2(±0.9) & 92.7(±1.1) \\
DA-RAG vs VanillaRAG & 93.9(±1.0) & 89.3(±1.8) & 89.9(±1.3) & 91.2(±2.1) & 87.2(±1.3) & 90.0(±1.1) & 91.3(±1.3) & 91.7(±1.7) & 94.8(±1.6) & 90.4(±1.2) & 90.1(±0.9) & 91.6(±2.1) \\
\midrule
\rowcolor[HTML]{F2F2F2}
\multicolumn{13}{l}{\textbf{Graph-based RAG}} \\
DA-RAG vs GLightRAG & 90.8(±0.7) & 90.1(±1.0) & 91.4(±1.0) & 91.1(±0.9) & 85.5(±0.4) & 92.9(±0.7) & 93.1(±0.4) & 93.7(±0.9) & 94.4(±0.7) & 93.1(±0.9) & 93.2(±0.4) & 93.1(±0.9) \\
DA-RAG vs HLightRAG & 90.7(±0.4) & 89.7(±0.9) & 90.0(±1.1) & 89.5(±1.7) & 85.3(±0.8) & 88.9(±0.8) & 90.7(±0.4) & 90.6(±0.4) & 92.9(±0.7) & 88.7(±0.7) & 92.1(±1.1) & 91.1(±0.4) \\
DA-RAG vs MLightRAG & 90.0(±1.6) & 83.3(±2.3) & 87.1(±2.4) & 87.0(±1.5) & 85.7(±0.4) & 87.7(±1.3) & 90.1(±1.7) & 89.9(±2.0) & 91.6(±0.4) & 89.2(±1.1) & 92.1(±0.9) & 90.2(±1.0) \\
DA-RAG vs RAPTOR & 84.3(±1.1) & 77.2(±2.2) & 82.5(±2.3) & 81.2(±1.0) & 88.7(±0.7) & 75.5(±1.9) & 86.4(±0.7) & 87.3(±0.7) & 86.2(±1.3) & 80.8(±0.8) & 85.5(±1.2) & 84.3(±0.6) \\
DA-RAG vs HippoRAG & 82.2(±2.7) & 74.3(±1.6) & 77.5(±2.2) & 76.4(±1.1) & 89.4(±1.2) & 73.7(±0.7) & 82.3(±1.9) & 82.3(±2.3) & 82.2(±1.5) & 84.4(±1.1) & 88.2(±1.3) & 86.9(±1.9) \\
DA-RAG vs LGraphRAG & 70.9(±1.3) & 67.0(±1.6) & 69.2(±1.6) & 67.8(±2.4) & 87.6(±1.7) & 80.8(±3.4) & 84.9(±1.2) & 83.3(±2.4) & 77.0(±1.6) & 75.7(±1.6) & 72.3(±1.0) & 75.3(±1.1) \\
DA-RAG vs GGraphRAG & 57.8(±2.9) & 57.1(±3.2) & 59.1(±3.9) & 56.6(±4.2) & 60.7(±3.4) & 50.7(±1.0) & 57.7(±3.8) & 55.2(±1.9) & 59.9(±1.5) & 61.5(±1.3) & 63.5(±2.1) & 61.7(±0.8) \\
DA-RAG vs ArchRAG & 50.3(±3.9) & 59.5(±2.7) & 53.1(±4.3) & 55.7(±2.9) & 52.6(±1.9) & 53.9(±2.1) & 58.2(±3.3) & 52.1(±3.6) & 52.4(±3.9) & 58.7(±2.4) & 55.5(±2.5) & 56.9(±2.6) \\
\bottomrule
\end{tabular}}
\label{tab:win_rates}
\end{table*}

\vspace{-8pt}
\section{Experiments}\label{sec:evaluation}

% In this section, we conduct a thorough experiment to evaluate the performance of DA-RAG. 
In this section, we conduct a thorough experiment to evaluate the performance, answering the following research questions (RQs):
\begin{itemize}
    \item \textbf{RQ1}: How does DA-RAG perform compared to baselines?
    \item \textbf{RQ2}: How efficient is the DA-RAG approach?
    \item \textbf{RQ3}: How is community quality retrieved by the DA-RAG method, and how does it affect RAG's performance?
\end{itemize}

\vspace{-10pt}
\subsection{Experimental Settings}
% \libin{\textbf{All the prompts to LLM for either generating input questions or evaluating results or generate community report or built sematic chunk node as mentioned below can be found in Appendix B.}}
% All prompts for generating input questions, evaluating results, creating community reports, or building semantic chunk nodes are detailed in Appendix B.
\textit{\textbf{Datasets}.} Specifically, we utilize the \textbf{Agriculture} and \textbf{Mixed} subsets from the UltraDomain benchmark \cite{Ultradomain}. We also include the \textbf{News Articles} \cite{MultiHop} dataset, previously employed in evaluating Microsoft's GraphRAG \cite{GraphRAG}. %Detailed statistics of these datasets are presented in Table \ref{tab:dataset_stats}. 
We prompted the LLM to generate 125 challenging questions for each dataset following~\cite{lightRaG} for comprehensive evaluation.

\textit{\textbf{Evaluation Metrics}.}
% Following earlier researches \cite{lightRaG}, we employ a head-to-head comparison approach using LLM \cite{head2head} as the evaluator. Recognizing that the presentation order of answers can introduce \textbf{positional bias} in LLM-based pairwise comparisons\cite{bias}, 
We follow the studies \cite{bias} to try both position orders ([$R_a, R_b$] and $[R_b, R_a]$) for each pair of evaluated RAG responses, and report the average win rate over all the questions and both position settings. 
% The LLM evaluator was prompted to provide detailed textual justifications for its preference over the responses from the two compared RAG approaches.
There are four evaluation dimensions consistent with recent RAG studies \cite{lightRaG,In-depthGrag}: \textbf{Comprehensiveness, Diversity, Empowerment, and Overall}. 
Please see Appendix~\ref{app: metrics} for more details on metrics.
% as follows:
% \begin{itemize}
%     \item \textbf{Comprehensiveness:} How much detail does the answer provide to cover all aspects and details of the question?
%     \item \textbf{Diversity:} How varied and rich is the answer in providing different perspectives and insights on the question?
%     \item \textbf{Empowerment:} How well does the answer help the reader understand and make informed judgments about the topic?
%     \item \textbf{Overall:} This dimension assesses the cumulative performance across the three preceding criteria to identify the best overall answer.
% \end{itemize}

\textit{\textbf{Baselines}.}
We evaluate our proposed method against several key baseline approaches from various representative G-RAG strategies. These include \textbf{LightRAG} \cite{lightRaG}, \textbf{HippoRAG} \cite{hippoRAG}, \textbf{RAPTOR} \cite{RAPTOR}, and community-partition-based methods such as \textbf{ArchRAG} \cite{ArchRAG} and \textbf{Microsoft's GraphRAG} \cite{GraphRAG}. Additionally, we incorporate \textbf{VanillaRAG} \cite{rag2020nuerips} and \textbf{BM25} \cite{BM25} as fundamental baselines. Our evaluation also considers the use of an LLM for answering questions without retrieval, specifically in \textbf{Zero-Shot} and \textbf{CoT} \cite{Cot} contexts.
All implementation details can be found in the Appendix~\ref{app: B.1}.
 %All baseline models were configured using their default hyperparameter settings to ensure a fair comparison.
% and direct question answering without any retrieval data, i.e., \textbf{Zero-Shot} 

%We compare our proposed \modelname{} against MicroSoft's GraphRAG, ArchRAG, LightRAG, HippoRAG,RAPTOR baselines and a Vanilla RAG approach.  Particularly, GraphRAG and LightRAG have several versions; we opt for \textbf{GraphRAG-Global},\textbf{LightRAG-Hybrid}, recognized as their strongest variant.

% \textit{\textbf{Implementation Details}.}
% Our experiments primarily utilize OpenAI's \textit{GPT-4o-mini} as the backbone LLM for response generation and \textit{text-embedding-3-small} for generating text embeddings. To reduce the randomness caused by the LLM, we set the response temperature to 0. For constructing the similarity layer within our chunk-layer oriented graph index, we employ $k$-Nearest Neighbors (KNN) to create edges between entities, with $k_{neighbor}$ set to 5. 

% Our code is available at \url{https://github.com/zxyangyu/DA-RAG}.
% \textbf{We upload our code to the supplementary material, where an appendix also describes more implementation details including the tiktoken model, embedding dimensions, entity gleaning times, and the context budget, among others.}
%\textbf{Our code and implementation details are available in the supplementary materials.} %, including prompts for question generation, evaluation, report creation, and semantic node building, along with configurations for the tiktoken model, entity gleaning times, and context budget.}

% Our code and data are available at \url{https://anonymous.4open.science/r/DA-RAG-BBF5/}.

\begin{figure}
    \centering
    \includegraphics[width=1\linewidth]{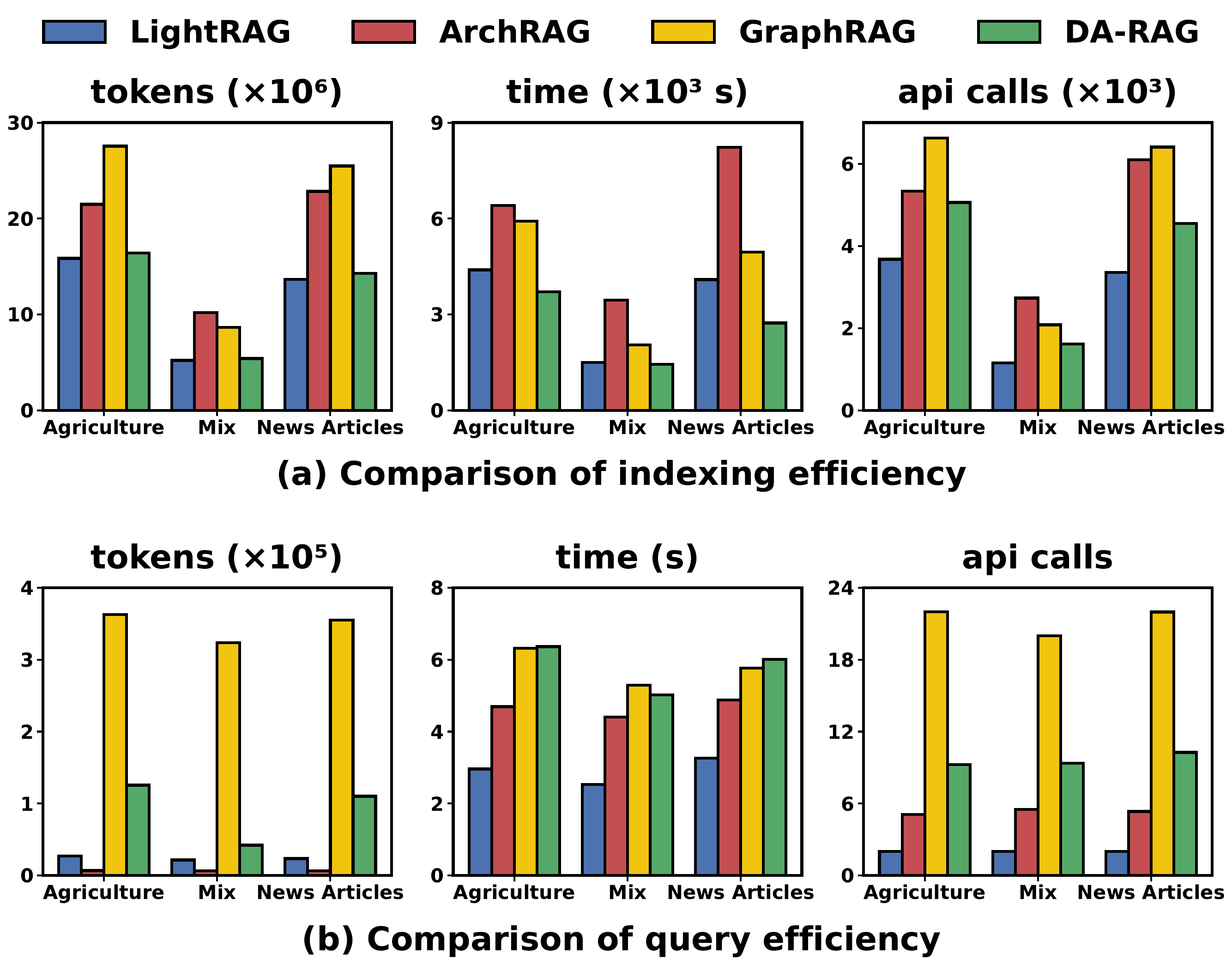}
        \vspace{-15pt}
    \caption{Efficiency comparison. 
    DA-RAG is more efficient than ArchRAG and GraphRAG, while surpassing LightRAG in terms of effectiveness when given comparable efficiency.
    }
    \vspace{-9pt}
    \label{fig: eff}
\end{figure}

\vspace{-5pt}
\subsection{Main Results}
% For main evaluations, we analyze \textit{effectiveness} and \textit{efficiency}:

\textit{\textbf{Effectiveness (RQ1)}.}
\label{sec: Main Results}
We conducted head-to-head comparisons of \modelname{} against all baseline models across four evaluation dimensions for each dataset. The win rates are presented in Table~\ref{tab:win_rates}, where each row represents one baseline. \textbf{\modelname{} consistently and significantly outperforms all baselines across all datasets and evaluation dimensions.}

% Notably, the substantial margins against VanillaRAG and LightRAG (for example, achieving average overall win rates of 91.5\% and 92.6\%, respectively) highlight the advantages of utilizing higher-order graph structures. The performance improvements are especially pronounced in the Comprehensiveness and Empowerment dimensions. Compared to Microsoft's GraphRAG-Global, which uses static graph communities, \modelname{} exhibits superior performance, achieving average win rates of 59.5\% in Comprehensiveness and 60.1\% in Empowerment. This indicates that \modelname{}'s dynamic, query-adaptive community retrieval, enabled by the chunk-layer oriented graph index, allows for a more thorough and contextually relevant synthesis of information, thereby empowering users to tackle complex tasks effectively. 

The complexity of the baseline models can stratify the analysis. First, the overwhelming win rates against non-retrieval methods (Zero-shot, CoT) and standard retrieval baselines (BM25, VanillaRAG) confirm the fundamental value of the RAG paradigm. For instance, the average overall win rate of 91.5\% against VanillaRAG underscores the inherent limitations of simple dense vector retrieval and highlights the initial benefits of using structural information.

Second, when compared to Graph-based RAG methods that leverage low-order graph structures, confined to immediate
neighbors or connected paths, \modelname{} maintains a commanding lead. Notably, the substantial margins against the LightRAG variants (e.g., an average overall win rate of 92.6\%) showcase the advantages of exploiting higher-order graph information (community). 

Finally, the comparison with more advanced models, such as Microsoft's GraphRAG and ArchRAG, is most revealing. As hypothesized, \modelname{}'s architecture demonstrates a clear advantage. The average win rates of 59.5\% in Comprehensiveness and 60.1\% in Empowerment against GGraphRAG validate the superiority of our dynamic, on-the-fly community identification over static, pre-partitioned graph communities. Our chunk-layer oriented graph index enables this query-specific context formation, leading to more thorough and relevant information synthesis.

\textit{\textbf{Efficiency (RQ2)}.}
We assessed the efficiency of \modelname{} by measuring both its time cost (in seconds) and token usage (for LLM calls). 
% This was done in comparison to GraphRAG and ArchRAG, which are both community-oriented approaches similar to DA-RAG. We also included LightRAG as a lightweight and efficient baseline for comparison.
As illustrated in Figure~\ref{fig: eff}(a), the construction of the index reveals significant advantages for \modelname{} when compared to community-partition-based approaches. Peak reductions reach 37.3\% in time and 41.8\% in tokens with larger datasets like News Articles (when compared to GraphRAG).   \textbf{\modelname{} consistently achieves optimal or near-optimal results} in terms of index construction efficiency.
% While the construction overhead of \modelname{} is comparable to that of simpler, non-community-focused baselines like LightRAG, it significantly outperforms other community-partition-based methods, such as GraphRAG and ArchRAG, in effectiveness. This demonstrates an excellent balance between advanced graph-based retrieval capabilities and computational cost.
% In online querying, as illustrated in Figure~\ref{fig: eff}(b), \modelname{} while maintaining a retrieval time comparable to that of GraphRAG, it achieves substantial reductions in total token usage. Specifically, \modelname{} cuts total token consumption by an average of 73.8\% (reaching up to 88.76\% in the Mix dataset) when compared to GraphRAG-Global. This significant efficiency gain is directly attributable to our proposed coarse-to-fine retrieval strategy.
In online querying, as illustrated in Figure~\ref{fig: eff}(b), \modelname{} maintains retrieval latency comparable to GraphRAG-Global while cutting total token consumption by an average of 73.8\% (up to 88.76\% on the Mix dataset). Specifically, our method imposes a total burden of only 9.3 API calls (42K tokens), including 8.3 calls (30K tokens) for the adaptive $k$-determination, which remains far below GraphRAG’s average of 21.3 calls (323K tokens). This significant efficiency gain is directly attributable to our proposed coarse-to-fine retrieval strategy.

\textbf{To summarize, DA-RAG outperforms all the baselines while maintaining satisfactory running efficiency.}

\subsection{Further Analysis (RQ3)}
\label{sec:Further Analysis}
\begin{figure}[t]
    \centering
    \includegraphics[width=1\linewidth]{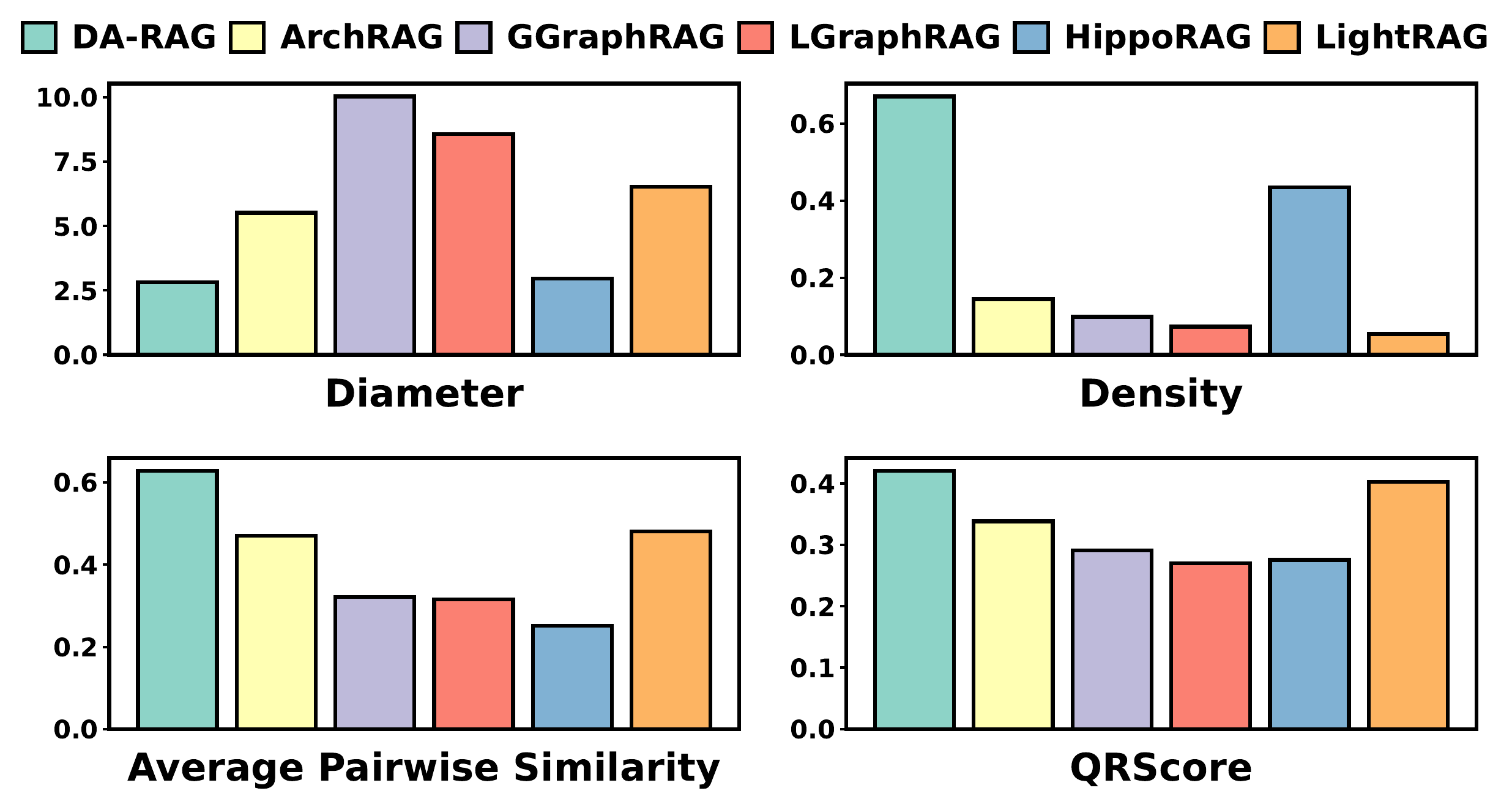}
    \caption{Analysis of retrieved subgraph quality on the Agriculture dataset. DA-RAG's subgraphs exhibit the superior structural cohesiveness (highest density and lowest diameter) and semantic relevance (highest QRScore and similarity)  compared to all baseline methods.
    }
    \label{fig:feature_statistics_comparison}   \Description{}

\end{figure}

\add{\vspace{3pt}\textit{\textbf{Comparative Analysis of Subgraph Quality}.}
\label{par:Analysis of Retrieved Subgraph Properties}
To provide deeper insight into the effectiveness of our approach, we analyze the properties of the subgraphs retrieved by DA-RAG and baseline methods. As illustrated in \autoref{fig:feature_statistics_comparison} for the Agriculture dataset, we assess four key metrics: diameter, density, QRScore and average pairwise similarity 
(detailed definitions can be found in the Appendix~\ref{sec:appendix_metrics}). 
The results unequivocally demonstrate the superior quality of the subgraphs identified by DA-RAG.

\textbf{Structurally, DA-RAG produces subgraphs with high density and low diameter.} These properties are direct consequences of our EACS formulation, which leverages the k-truss constraint to ensure \textbf{Provable Cohesion} and \textbf{Bounded Reasoning Hops}. In contrast, low-order retrieval methods (e.g., HippoRAG, LightRAG) yield sparse and disconnected contexts. \textbf{Semantically, DA-RAG achieves the retrieval of both highly relevant to the query and internally coherent topics} as indicated by the highest QRScore and average pairwise node similarity
%Semantically, DA-RAG achieves the highest QRScore and average pairwise node similarity, indicating that the retrieved subgraphs are both highly relevant to the query and internally coherent in topic. 
Moreover, while static community-based methods like GraphRAG can produce dense clusters, their lower semantic scores highlight a critical weakness: pre-computed partitions cannot adapt to the specific focus of a dynamic query.

\textbf{In summary, it confirms that DA-RAG excels at retrieving context that is both semantically focused and structurally cohesive.} This ability to construct a high-quality, compact knowledge subgraph for the LLM contributes to the significant performance gains observed in our main results (Section~\ref{sec: Main Results}).
}

\vspace{3pt}\textit{\textbf{Case Study}.}
We present a detailed case study in Table~\ref{tab:case_study}. The table contrasts the final contexts constructed by our DA-RAG framework and the GraphRAG baseline for the identical query: \textit{``How does celebrity endorsement shift consumer purchasing decisions?''} The results reveal a stark difference in context quality. The context generated by DA-RAG is highly coherent and directly addresses the query. It successfully retrieves not only a real-world example, \textbf{``Taylor Swift effect''} on NFL merchandise sales, but also a \textbf{general principle of celebrity influence}. In contrast, the context produced by GraphRAG suffers from rigid subgraph partitioning, resulting in an overload of irrelevant information that does not relate to the query's intent. This case vividly illustrates the core problem we identified in our introduction: the rigidity of the subgraph partition introduces substantial noise. 

\begin{table}[t!]
\caption{Case study comparing context retrieved by DA-RAG and GraphRAG. The context is abridged for clarity. }
\label{tab:case_study}
\centering
\small
\begin{tabular}{@{}p{1\columnwidth}@{}} % Using a single 'p' column that spans most of the column width
\toprule
\textbf{Query:} \textit{How does celebrity endorsement shift consumer purchasing decisions?} \\
\midrule
\textbf{DA-RAG Context:}\\
\textbf{Taylor Swift effect:} \dots Following the publicization of their relationship... Kelce's jersey sales soared by 400\% 
% illustrating how the association with Swift has incentivized her fanbase to engage with NFL merchandise 
\dots \\
\textbf{Celebrity Endorsements and Brand Visibility:} Products benefit from celebrities who use them. \dots This endorsement amplifies brand visibility and creates aspirational value \dots\\
\textbf{Engagement of Fanbase:} The engagement of Swift's extensive fanbase has drawn in millions, notably increasing ratings for games \dots \\
\midrule
\textbf{GraphRAG Context:} \\
\textbf{Celebrity endorsements and their fallout: }\dots
However, this association has since resulted in reputational damage for these individuals as FTX's practices come under intense scrutiny \dots \\
\textbf{The role of Celebrity Endorsements: }
Celebrity Endorsements are strategically leveraged to influence Gen Z's political perceptions \dots The marketing approach by FTX underlined the importance of these endorsements.  \dots\\

\bottomrule
\end{tabular}
\end{table}

\vspace{3pt}\textit{\textbf{Ablation Study}.}
To dissect the contributions of the core components of \modelname{}, we compare \modelname{} with the following variants:

\begin{itemize}
\item \textbf{\modelname{} w/o Similarity Layer (w/o $L_S$)}: This variant removes the Similarity Layer ($L_S$) from the graph index. 
\item \textbf{\modelname{} w/o Semantic Chunk Layer (w/o $L_C$)}:  The EACS process is performed directly on the Knowledge Graph Layer ($L_{KG}$) and Similarity Layer ($L_S$).% In this configuration, the Semantic Chunk Layer ($L_C$) is omitted.
    \item \textbf{\modelname{} w/o Semantic Chunking (w/o SC)}: Replaces the semantic chunking method with a fixed-size chunking approach (1200 tokens/chunk with a 100-token overlap).
    \item \textbf{\modelname{} w/o ACS (using 1-hop Retrieval)}: Replaces the EACS module with a widely adopted retrieval strategy \cite{DALK}. 
\end{itemize}

Table~\ref{tab:detail_ana}(a)
illustrates the individual contributions of key components in \modelname{}. 
% Each unit presents the win rate of the captioning method on the left compared to other methods.
\textbf{Our analysis shows that removing any single component leads to a noticeable decline in performance across all metrics.} In particular, excluding our dynamic attributed community search method (w/o ACS) or the Semantic Chunk Layer (w/o $L_C$) results in the most significant drops in performance. Specifically, \modelname{} w/o EACS shows win rates ranging from 25\% to 41\% compared to \modelname{} across four metrics. Similarly, \modelname{} w/o $L_C$ experiences a decline in win rates from 50\% to between 21\% and 34\%. These substantial decreases emphasize the critical role of EACS in retrieving relevant and cohesive subgraphs, while the $L_C$ layer facilitates effective access to information at multiple granularities.

\vspace{3pt} \textit{\textbf{Varying LLMs}}.
We evaluated our framework's sensitivity to the choice of the LLMs, with results summarized in Table~\ref{tab:detail_ana}(b). This table presents the win rates of our DA-RAG method compared to Microsoft’s GraphRAG-Global baseline. \textbf{We found that the performance of DA-RAG improves when using more powerful integrated language models.} We posit that more advanced models like \textit{gpt-4o} provide a richer and more accurate information substrate for our index. This enhancement amplifies the effectiveness of our DA-RAG framework by enabling adaptive community search to identify more coherent and semantically rich subgraphs tailored to specific queries, resulting in superior outcomes.

\begin{table}[t]

\caption{(a) Win rates of DA-RAG variants vs. full model. (b) Sensitivity analysis on different LLM backbones.}
\label{tab:detail_ana}
\centering
\small
\begin{tabularx}{\linewidth}{lXXXX}
\toprule
\textbf{Configuration} & \textbf{Comp.} & \textbf{Div.} & \textbf{Emp.} & \textbf{Over.} \\
\midrule
% 第二个子表：消融实验
\multicolumn{5}{l}{\textit{\textbf{(a) Ablation Study}}} \\[4pt] % 子标题，跨越5列，左对齐
\hspace{1em} w/o ACS       & 25.25\% & 30.56\% & 41.90\% & 33.33\% \\[3pt]
\hspace{1em} w/o $L_C$     & 21.43\% & 34.95\% & 25.53\% & 32.20\% \\[3pt]
\hspace{1em} w/o SC        & 40.21\% & 30.00\% & 31.96\% & 35.65\% \\[3pt]
\hspace{1em} w/o $L_S$     & 44.68\% & 40.78\% & 39.78\% & 40.52\% \\
\midrule
% 第一个子表：不同LLM backbone的对比
\multicolumn{5}{l}{\small \textit{\textbf{(b) Varying LLMs}}} \\[4pt] % 子标题，跨越5列，左对齐
\hspace{1em} gpt-3.5-turbo & 57.54\% & 59.57\% & 60.78\% & 58.36\% \\[3pt]
\hspace{1em} gpt-4o-mini   & 59.62\% & 61.74\% & 58.42\% & 58.12\% \\[3pt]
\hspace{1em} gpt-4o        & 60.43\% & 63.44\% & 62.38\% & 62.24\% \\
\bottomrule
\end{tabularx}%
\end{table}

\vspace{3pt}\textit{\textbf{Hyperparameter Sensitivity Analysis}}
We analyzed the sensitivity of our DA-RAG model to the hyperparameter $k$, the number of nearest neighbors used to construct the Similarity Layer ($L_s$). As shown in Figure~\ref{fig:pk}, the model demonstrates strong robustness, with its win rates against Microsoft's GraphRAG fluctuating only minimally as $k_{neighbor}$ varies from 3 to 9.
We attribute this stability to our multi-perspective design, where the final context is synthesized from several layers ($L_C$, $L_{KG}$, and $L_S$). In this framework, the $L_S$ layer acts as a complementary viewpoint by capturing essential node similarities, rather than being the sole driver of performance. Given that $k_{neighbor}=5$ offers a consistent advantage, it is set as the default value in our experiments.

\add{    \vspace{-10pt}
\section{Related Work}

Existing graph-based RAG approaches can be broadly categorized by how deeply they exploit structural information in graph data: \textbf{Adjacency Retrieval}, \textbf{Graph Topology Awareness}, and \textbf{High-order Pattern Mining}.

\textbf{Adjacency Retrieval.}
Methods in this category~\cite{lightRaG,GRAG-NAACL,ToG,tog2,DALK,MindMap,SubgraphRetrievalEnhancedModel,SimpleQuestion,SimpleisEffective} utilize low-order structural signals, typically confined to immediate neighbors or connected paths.
For example, LightRAG~\cite{lightRaG} retrieves top-$k$ relevant entities and relations from the embedding space, followed by a one-hop expansion to create a subgraph as context. 
In addition, other works~\cite{ToG,tog2,DALK,MindMap,SubgraphRetrievalEnhancedModel} employ Large Language Models (LLMs) to iteratively traverse graphs, exploring neighborhoods and constructing reasoning paths for inference.
Nevertheless, all these methods are constrained by their local perspective, often failing to capture the broader, global associations present within the graph.

\textbf{Graph Topology Awareness.}
To move beyond local adjacency, some research incorporates graph algorithms to capture the structural importance of nodes. 
HippoRAG~\cite{hippoRAG,hipporag2} applies PageRank~\cite{page1999pagerank} to assign global relevance scores to nodes given a query. 
Other works ~\cite{GNN-RAG,RGNN-Ret,KagNet,DualReasoning,GreaseLM,QAGNN,GrapeQA} train Graph Neural Networks (GNNs)~\cite{GNN} to score node relevance, extracting top nodes and their connecting paths as context.
Although these techniques enhance structural awareness, their focus typically remains on scoring individual nodes or simple paths. Consequently, they often overlook semantically richer, high-order patterns, such as communities or thematic clusters, which represent more abstract concepts.

\textbf{High-order Pattern Mining.}
Recent work highlights the value of high-order structures in knowledge graphs, including communities, cliques, and other meaningful subgraph patterns. 
RAPTOR~\cite{RAPTOR} organizes text into hierarchical trees via recursive clustering, capturing higher-level semantic groupings.
Likewise, methods ~\cite{GraphRAG,ArchRAG,communityRAG} detect community structures and produce summary reports for each, demonstrating notable gains in query-focused summarization tasks. However, most advanced methods \cite{GraphRAG,ArchRAG} rely on static, pre-computed structures that may not align with specific queries. To address this, we propose a framework that can dynamically discover structurally cohesive and semantically relevant subgraphs on-the-fly, tailored specifically to each query.
\begin{figure}[!t]
    \centering
    \includegraphics[width=1\linewidth]{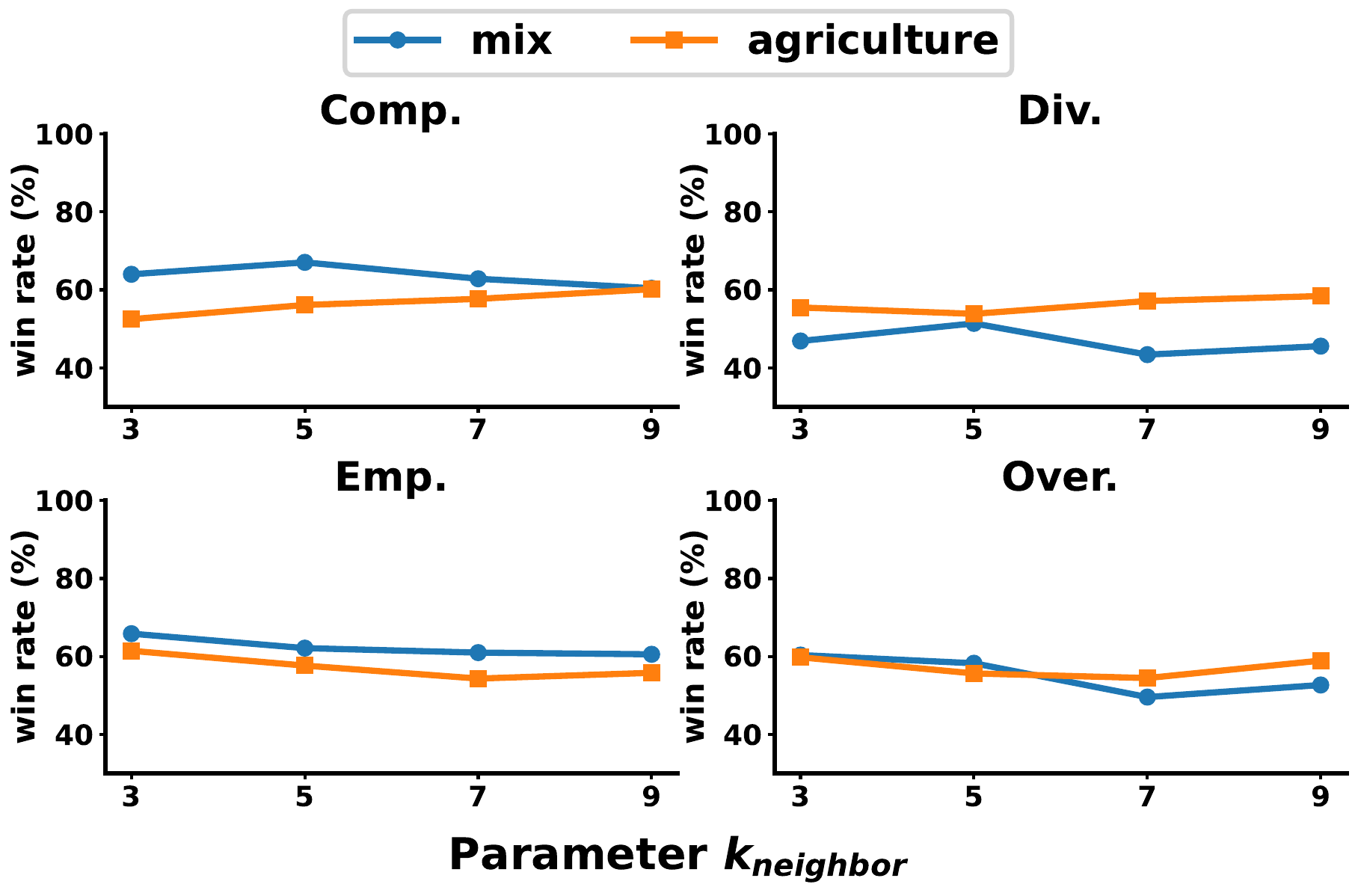}
    \caption{Sensitivity Analysis on $k_{neighbor}$, where DA-RAG is considerably stable and $k_{neighbor}=5$ offers the best results.}
    \label{fig:pk}   \Description{}
    \vspace{-5pt}
\end{figure}}

    \vspace{-5pt}
\section{Conclusion}
\label{sec:5}
% \modelname{} effectively leverages graph data structures, significantly enhancing retrieval accuracy and reducing token overhead within G-RAG framework. Our findings highlight the considerable potential of G-RAG to improve LLM augmentation and advance real-world applications.

% Despite these contributions, \modelname{} presents limitations that define avenues for future work. Its current optimization for AbstractQA \cite{In-depthGrag}, following precedents like Microsoft's GraphRAG \cite{GraphRAG} and LightRAG \cite{lightRaG}, calls for the development of adaptive retrieval strategies for diverse tasks such as SpecificQA \cite{In-depthGrag}. Furthermore, potential performance variability due to external API dependencies underscores the need for more controlled evaluation environments in subsequent studies. We acknowledge the dual societal impacts of RAG technologies; a comprehensive discussion on these aspects and potential mitigation strategies is provided in Appendix C.

In this paper, we addressed the limitations of coarse-grained topological exploration in existing G-RAG methods, i.e., low-order structural information and static community partitions are inadequate for handling dynamic user queries. We introduced \modelname{}, which adaptively identifies and retrieves relevant knowledge subgraphs based on the semantics of the queries. Our experiments demonstrated the superior effectiveness of \modelname{}, achieving an average win rate of 57.34\% over the GraphRAG-Global baseline, thus validating the benefits of dynamic retrieval. This work bridges graph analytics and retrieval-augmented generation, setting a new performance benchmark and establishing a novel methodological path.

%This fusion of disciplines is crucial for unlocking the next level of intelligence and reasoning capabilities in large-scale AI.
%Furthermore, its chunk-layer oriented index design reduces construction time and token overhead by up to 37\% and 41\%, respectively. 

\vspace{-5pt}
\begin{acks}
This work is supported by the National Natural Science Foundation of China (No. 62472455, U22B2060), Key-Area Research and Development Program of Guangdong Province (2024B0101050005),  Research Foundation of Science and Technology Plan Project of  Guangzhou City (2023B01J0001, 2024B01W0004). Chen Zhang is supported by the NSFC/RGC Joint Research Scheme sponsored by the Research Grants Council of Hong Kong and the National Natural Science Foundation of China (Project No. N\_PolyU5179/25); 
2) the Research Grants Council of the Hong Kong Special Administrative Region, China (Project No. PolyU25600624); 
3) the Innovation Technology Fund (Project No. ITS/052/23MX and PRP/009/22FX).
\end{acks}
\newpage
\bibliographystyle{ACM-Reference-Format}
\balance
\bibliography{references}

\newpage
\appendix
\section{Problem Analysis}
The $k$-truss \cite{trusses} is an important cohesive subgraph concept extensively studied in the literature \cite{densitybound, TrussDecomposition, ATC}. Burkhardt explored edge count ranges \cite{densitybound} for the connected $k$-truss, while Cohen established minimum diameter bounds \cite{diameter}. In Appendices \ref{ap:a1} and \ref{ap:a2}, we provide concise alternative proofs for the density and diameter bounds of the connected $k$-truss. Next, we formally define the free-rider effect for the G-RAG in Appendix \ref{ap:a3} and demonstrate that our proposed EACS problem effectively mitigates this effect. Lastly, Appendix \ref{ap:a4} contains complexity hardness proofs for EACS and an analysis of Algorithm 1 in Appendix \ref{ap:a5}.

\subsection{Lower Bound on Density of Connected $k$-truss}
\label{ap:a1}
While Burkhardt \cite{densitybound} provided a range of edge counts for the connected $k$-truss, the density bound was not investigated further. 
In this part, we establish a density bound from a different perspective based on vertex degrees. Although the proof of the density bound is relatively trivial, to the best of our knowledge, it has not been formally provided previously.

\textbf{Theorem}.
The density of a $k$-truss with $|V|$ vertices is greater than $\frac{k-1}{|V|-1}$.

\textbf{Proof}.
First, recall the definition of a $k$-truss:
\begin{itemize}
    \item In a $k$-truss, each edge must participate in at least $(k-2)$ triangles.
    \item This implies that the two endpoints of each edge must have at least $(k-2)$ common neighbors.
\end{itemize}
According to the degree sum formula in graph theory:
\begin{equation}
    \sum_{v \in V} deg(v) = 2|E|.
\end{equation}

Since each vertex has degree at least $(k-1)$, we have:
\begin{equation}
    2|E| = \sum_{v \in V} deg(v) \geq |V|(k-1).
\end{equation}

Therefore:
\begin{equation}
    |E| \geq \frac{|V|(k-1)}{2}.
\end{equation}

The density of a graph is defined as:
\begin{equation}
    density = \frac{2|E|}{|V|(|V|-1)}.
\end{equation}

Substituting the inequality from above:
\begin{equation}
    density \geq \frac{|V|(k-1)}{|V|(|V|-1)} = \frac{k-1}{|V|-1}.
\end{equation}

Thus, we have proven that the density of a $k$-truss is indeed greater than $\frac{k-1}{|V|-1}$.

\subsection{Upper Bound on Diameter of Connected $k$-truss}
\label{ap:a2}
This result has already been stated by Jonathan Cohen \cite{trusses}. Here we present a novel proof approach based on Breadth-First Search (BFS) spanning tree. 

\textbf{Theorem}.
For any connected $k$-truss with $|V|$ vertices, its diameter $d$ is at most $\lfloor \frac{2\,|V|-2}{k} \rfloor$.

\textbf{Proof}.
Without loss of generality, select a vertex $r \in V$ and construct a breadth-first search (BFS) spanning tree $T$ rooted at $r$ such that the height $t$ of $T$ is maximized. By definition, the diameter satisfies $d =t$.

We make two key observations regarding the BFS tree $T$:
\begin{enumerate}
    \item Every edge of $T$ corresponds to an edge in the original graph $H$.
    \item No edge in $H$ connects a vertex at level $i$ in $T$ to a vertex at level $i+2$ or higher, due to the BFS layer structure.
\end{enumerate}

Consider an edge $(u,v)$ in the BFS tree between $L_i$ and $L_{i+1}$. Since $H$ is a $k$-truss, the two endpoints of edge $(u,v)$ must have at least $(k-2)$ common neighbors. The common neighbors must lie within levels $i$ or $i+1$ due to observation 2. Therefore, for each $i=0,1,\dots,t-1$ the total number of vertices in the two consecutive layers $|L_i|$, $|L_{i+1}|$ satisfies
\begin{equation}
        |L_i| + |L_{i+1}| \ge (\underbrace{1+1}_{|u|+|v|}) + (k-2) = k.
\end{equation}

Summing this inequality over $i=0,1,\dots,t-1$ gives
\begin{equation}
\begin{split}
    \sum_{i=0}^{t-1}\bigl(|L_i| + |L_{i+1}|\bigr)
    &= \bigl(|L_0| + |L_t|\bigr) + 2 \sum_{i=1}^{t-1} |L_i| \\
    &= 2\,|V(H)| - \bigl(|L_0| + |L_t|\bigr) \\
    &\ge t\,k.
\end{split}
\end{equation}

Finally, since $|L_0|=1$, and in the worst case minimizing total vertices we also take $|L_t|=1$, we obtain
\begin{equation}
    2\,|V| \;- 2 
    \;\ge
    t\,k
    \quad\Longrightarrow\quad
    t \;\le\; \frac{2\,|V| - 2}{k}.
\end{equation}
As $d=t$ is an integer, the result follows:
\begin{equation}
    d \;\le\; \left\lfloor \frac{2\,|V| - 2}{k} \right\rfloor.
\end{equation}

\subsection{Mitigates the Free-rider Effect}
\label{ap:a3}
\textbf{Definition (Free-rider Effect for RAG)}.
Given a query relevance score $f(q,\cdot)$ (the larger, the better). 
Let $H(V_h,E_h,A_h)$ be an optimal solution of the subgraph retrieval problem within the graph $G(V,E,A)$ for the query $q$. 
If there exists a node set \( S \subseteq V \), where \( S \not\subseteq V_h \), 
such that 
\[f(q,G[S \cup V_h]) \geq f(q,H), \]
where \( G[S \cup V_h] \) denotes the subgraph induced by node set \(S \cup V_h\), 
then we say that the subgraph retrieval problem suffers from the free rider effect based on $f(q,\cdot)$.

\textbf{Theorem}.
Let $H$ be the discovered communities of the EACS with $q$.
For any subgraph node set \( S \subseteq V \),where \( S \not\subseteq V_h \), it holds that 
\[f(q,G[S \cup V_h]) < f(q,H).\]

\textbf{Proof}.
We prove this statement by contradiction. Assume, for the sake of contradiction, that there exists exists a node set \( S \subseteq V \),where \( S \not\subseteq V_h \),  such that
\(
f(q,H \cup H^*) \ge f(q,H).
\)

Since \( H\) is the solution to the EACS problem, it follows directly from its optimality that for any subgraph \( H' \subseteq G \), we have 
\begin{equation}
f(q,H') \le f(q,H).
\end{equation}

Combining this optimality property with our assumption, we have
\begin{equation}
f(q,H) \le f(q,G[S \cup V_h]) \le f(q,H).
\end{equation}

By the squeeze theorem, we then have equality:
\begin{equation}
    f(q,H) = f(q,G[S \cup V_h]).
\end{equation}

Now, since \( S \not\subseteq V_h \), it follows that
\begin{equation}
|G[S \cup V_h]| \ge |H|.
\end{equation}

We consider two possible cases separately:

Case 1: If \(G[S \cup V_h]\) is a connected \( k \)-truss, then \( G[S \cup V_h] \) itself could serve as a candidate community satisfying the constraints of the EACS problem with \( q \). \(|G[S \cup V_h]| \ge |H|\). This contradicts the maximality assumption of the discovered community \( H\), which is defined as an optimal solution to the EACS problem.

Case 2: If \( G[S \cup V_h]\) is not a connected \( k \)-truss, then it does not satisfy the constraints required by the EACS problem. Hence, it cannot qualify as a feasible solution to the EACS problem.

In both cases, we arrive at a contradiction. Therefore, our initial assumption that \( f(q,G[S \cup V_h]) \ge f(q,H) \) does not hold, and thus we must have
\begin{equation}
f(q,G[S \cup V_h]) < f(q,H).
\end{equation}

This completes the proof. Therefore, we conclude that the EACS problem can avoid the free rider effect.

% \textbf{EACS-Decision Problem.}. Given an high-dimensional attributed graph $G(V,E,A)$, natural language query $q$, parameters $k$, the EACS-Decision problem is testing whether there exists a connected subgraph $H \subseteq G$ as a $k$-truss such that $QRScore(H,q) \geq \delta$.

% \textbf{Theorem} The EACS-Decision problem is NP-Hard.

% \textbf{Proof}: We reduce a well-known NP-hard problem of Maximum Clique (decision version) to the EACS-Decision problem. Given a simple graph $G'=(V_{G'},E_{G'})$ and a parameter $k'$, the decision version of the Maximum Clique problem is to check whether $G'$ contains a clique of size $k'$.

% \textbf{Theorem} For any $\epsilon > 0$, it is hard to approximate the Embedding-Attributed Community Search problem in polynomial time within a factor of $n^{1-\epsilon}$. 

% \textbf{Proof}: 
% We prove this by reduction from the Maximum Clique problem, which is known to be inapproximable within a factor of $n^{1-\epsilon}$ for any $\epsilon > 0$ unless P = NP.

\subsection{Hardness}
\label{ap:a4}
In this section, we show the EACS is NP-hard. 
% We further show that EACS is also hard to approximate in any constant factor. 
To this end, we define the decision version of the EACS problem and first prove its decision problem is NP-hard.

\textbf{Problem (EACS-Decision)}. Given a attributed graph $G=(V,E,A)$, a natural language query $q$, parameter $k_t$, and threshold $\delta$, the EACS-Decision problem is to determine whether there exists a connected subgraph $H \subseteq G$ that is a $k_t$-truss and satisfies $QRScore(H, q) \geq \delta$. 

\textbf{Theorem}.
The EACS-Decision problem is NP-Hard.

\textbf{Proof}. We prove this by a polynomial-time reduction from the well-known NP-hard problem \textit{Maximum Clique} (decision version). Given an undirected graph $G' = (V', E')$ and an integer $k'$, the decision version of the Maximum Clique problem asks whether $G'$ contains a clique of size $k'$.

Given such an instance $(G', k')$, we construct an instance of the EACS-Decision problem as follows:

\begin{itemize}
    \item Let the attributed graph $G = (V, E, A)$ have the same graph structure as $G'$, i.e., $V = V'$, $E = E'$.
    \item For each node $v \in V$, assign the same attribute vector: $A_v = (1, 0, 0, \dots, 0)$.
    \item Define the query vector as $q = (1, 0, 0, \dots, 0)$.
    \item Set $k_t = k'$, and $\delta = 1$.
\end{itemize}

We now show that $G'$ contains a clique of size $k'$ if and only if the constructed EACS-Decision instance has a valid solution.

($\Rightarrow$) Suppose there exists a $k'$-clique $C$ in $G'$. Let $H$ be the induced subgraph of $G$ on the nodes in $C$.
\begin{itemize}
    \item $H$ is connected since a clique is fully connected.
    \item In a $k'$-clique, every edge is part of exactly $k'-2$ triangles. Thus, $H$ is a $k'$-truss (i.e., every edge is in at least $k'-2 = k_t-2$ triangles).
    \item Each node in $H$ has attribute vector $(1, 0, 0, \dots, 0)$, which matches the query vector, so $QRScore(H, q) = 1 \geq \delta$.
\end{itemize}
Therefore, $H$ is a valid solution to the EACS-Decision problem.

($\Leftarrow$) Suppose there exists a subgraph $H \subseteq G$ that is a connected $k_t$-truss and satisfies $QRScore(H, q) \geq \delta$.
\begin{itemize}
    \item Since $QRScore(H, q) \geq \delta = 1$, and all attribute vectors are $(1, 0, 0, \dots, 0)$.
    \item Since $H$ is a $k_t$-truss, each edge in $H$ participates in at least $k_t-2 = k'-2$ triangles.
    \item The only way to ensure this condition in general graphs (without creating additional structures) is to have $H$ be a $k'$-clique, since in a $k'$-clique each edge is in exactly $k'-2$ triangles.
\end{itemize}
Hence, $H$ corresponds to a clique of size $k'$ in $G'$.

The construction can be completed in polynomial time:
\begin{itemize}
    \item Copying the graph structure takes $O(|V| + |E|)$ time.
    \item Assigning attribute vectors takes $O(|V|)$ time.
    \item Setting parameters takes $O(1)$ time.
\end{itemize}

Therefore, EACS-Decision is NP-hard.

% \begin{algorithm}[b]
% \caption{Q-Peel (Query-aware Peeling)}
% \label{al:1}
% \begin{algorithmic}[1]
% \Require $G$: Undirected attributed graph, $q$: Query embedding, $k$: Target k for k-truss
% \Ensure $H$: Optimal community
% \State $T_k \gets$ maximal $k$-truss subgraph of $G$
% \State$C \gets$ connected components in $T_k$
% ; $updated \gets$ \textbf{true}
% \ForAll{component $c$ in $C$} $S \gets$ nodes in $c$ 
%   \While{updated} $updated \gets$ \textbf{false}
%     \ForAll{$v$ in \Call{SortByRelevance}{$S, q$}} 
%     \State $S' \gets S \setminus \{v\}$
%       \If{\Call{IsValidImprovement}{$S', S, q$}}
%         \State $S \gets S'$; $updated \gets$ \textbf{true}
%         \State \textbf{break}
%       \EndIf
%     \EndFor
%   \EndWhile
%   \If{\Call{QRScore}{$H', q$} $<$ \Call{QRScore}{$S, q$}}
%   \State $H' \gets S$
%  \EndIf
% \EndFor
% \State \Return $H \gets H'$
% \end{algorithmic}
% \end{algorithm}

\subsection{Complexity Analysis}
\label{ap:a5}

We now analyze the computational complexity of our proposed Algorithm, Q-Peel, with pseudocode in Algorithm~\ref {al:1}, focusing on both time and space aspects.

\begin{table*}[t]
\small
\centering
\caption{Detailed parameter configurations for the DA-RAG model.}
\setlength{\tabcolsep}{4pt}
\resizebox{\textwidth}{!}{
\begin{tabularx}{\linewidth}{l l X}
\toprule
\textbf{Parameter} & \textbf{Value} & \textbf{Description} \\
\midrule
language\_model\_name & gpt-4o-mini & Language model used for response generation. \\
embedding\_model\_name & text-embedding-3-small & Model used to generate vector embeddings. \\
evaluation\_model\_name & gpt-4o-mini & Language model used for head-to-head comparison. \\
tiktoken\_model\_name & gpt-4o & Model used for token counting and encoding. \\
entity\_extract\_max\_gleaning & 1 & Max iterations to refine entity extraction. \\
entity\_summary\_to\_max\_tokens & 500 & Max tokens allowed in the entity summary. \\
embedding\_dimensions & 1536 & Dimensionality of embedding vectors. \\
embedding\_max\_token\_size & 8192 & Maximum number of tokens that can be embedded at once. \\
embedding\_func\_max\_async & 16 & Max number of asynchronous embedding calls. \\
language\_model\_max\_async & 16 & Max number of asynchronous calls to the language model. \\
language\_model\_max\_token\_size & 32768 & Maximum context length supported by the language model. \\
key\_string\_value\_json\_storage & JsonKVStorage & Class used for JSON-based key-value storage. \\
vector\_db\_storage & NanoVectorDBStorage & Class managing vector database operations. \\
graph\_storage & NetworkXStorage & Class for graph-based storage using NetworkX. \\
max\_token\_for\_text\_unit & 4000 & Token budget for single text unit. \\
max\_token\_for\_context & 4800 & Token budget for retrieval context. \\
max\_token\_for\_community\_report & 3200 & Token budget for community report. \\
\bottomrule
\end{tabularx}
}
\label{tab:da_rag_parameters}
\end{table*}

\paragraph{Time Complexity}
The algorithm consists of several key stages. In Line~1, it computes the maximal $k$-truss subgraph of $G$. This can be done in $O(m^{1.5})$ time, where $m$ is the number of edges in the input graph $G$, using the standard $k$-truss decomposition algorithm~\cite{TrussDecomposition}. Next, the algorithm iterates over each connected component of the $k$-truss subgraph and applies a node refinement process (Lines~3--10). Let $c$ denote the number of such components, and $|S|$ denote the number of nodes in a component. During refinement, each node in $S$ is considered for removal in each iteration of the while-loop. In the worst case, there can be up to $O(|S|)$ iterations, with one node removed per iteration. The subroutine \textsc{IsValidImprovement} can be performed in $O(t)$ time, where $t$ is the number of edges in the 1-hop neighborhood subgraph of the removed node. Therefore, the total cost of processing one component is $O(|S|^2t)$. In the worst case, where $|S| = O(n)$, the cost becomes $O(n^2t)$. Considering all components, the worst-case time complexity of the algorithm is:
\[
O(m^{1.5} + cn^2t)
\]
where $c$ is the number of connected components in the $k$-truss, and $t$ is the upper bound on the size of 1-hop neighborhood subgraphs. In practice, both $c$ and $t$ are typically much smaller than $n$, and the actual runtime is significantly reduced due to early termination of the refinement process.
\paragraph{Space Complexity}
Let $n = |V|$ and $m = |E|$ be the number of nodes and edges in the input graph $G$, respectively. The algorithm maintains several auxiliary data structures:
\begin{itemize}
    \item The $k$-truss subgraph $T_k$, stored as a subset of $G$, requires $O(n + m)$ space.
    \item The list of connected components $C$ of $T_k$, requiring up to $O(n)$ space.
    \item For each component, a working node set $S$ and its variants $S'$, consuming $O(n)$ space per component.
    \item Temporary structures used during refinement, such as priority queues for \textsc{SortByRelevance}, boolean flags, and candidate communities, each requiring at most $O(n)$ space.
\end{itemize}
The most space-intensive operation is the maintenance of intermediate subgraphs during refinement. However, since these are all subgraphs of $T_k$, their cumulative space requirement remains bounded by $O(m)$.
Therefore, the overall space complexity of the algorithm is:
$O(n + m).$

\section{Experimental Details}
\label{app: B.1}
\subsection{Implementation Details}
Experiments were conducted on a Linux server equipped with an Intel Xeon 3.00 GHz CPU, 256 GB of RAM, and three NVIDIA GeForce RTX 3090 GPUs, each with 24 GB of VRAM. To reduce the randomness caused by the LLM, we set the response temperature to 0. For constructing the similarity layer within our chunk-layer oriented graph index, we employ $k$-Nearest Neighbors (KNN) to create edges between entities, with $k_{neighbor}$ set to 5. Detailed parameter configurations for the DA-RAG model can be found in Table \ref{tab:da_rag_parameters}. 
% For the baseline methods, we used publicly available implementations. 
% Among the versions of GraphRAG and LightRAG, we select GraphRAG-Local, GraphRAG-Global, LightRAG-Hybrid,LightRAG-Global and LightRAG-Mix for comprehensive comparisons. \par
For baseline implementation, we applied the code provided in DIGIMON \cite{In-depthGrag} for Hippo and RAPTOR. For GraphRAG, ArchRAG, and LightRAG, we utilized their officially released implementations.

\subsection{Evaluation Metrics}
\label{app: metrics}
To evaluate the quality of generated answers, we conduct a head-to-head comparison using an LLM-based evaluator. For this comparison, we adopt four metrics from previous work \cite{lightRaG,ArchRAG}, which are defined as follows.
 \textbf{Comprehensiveness:} How much detail does the answer provide to cover all aspects and details of 
    the question?
 \textbf{Diversity:} How varied and rich is the answer in providing different perspectives and insights on the question?
 \textbf{Empowerment:} How well does the answer help the reader understand and make informed judgments about the topic?
\textbf{Overall:} This dimension assesses the cumulative performance across the three preceding criteria to identify the best overall answer.

\subsection{Definitions of Subgraph Property Metrics}
\label{sec:appendix_metrics}

This section provides the formal definitions for the metrics used to evaluate the retrieved subgraphs. Let $\mathcal{Q}$ denote the set of evaluation queries. For each query $q \in \mathcal{Q}$, G-RAG approaches may retrieve a set of subgraphs, denoted as $\mathcal{H}_q$. The values reported in the main paper represent the average of each metric computed over the entire collection of retrieved subgraphs from all queries:
\[
\overline{\text{Metric}} = \frac{1}{\sum_{q' \in \mathcal{Q}} |\mathcal{H}_{q'}|} \sum_{q \in \mathcal{Q}} \sum_{H \in \mathcal{H}_q} \text{Metric}(H, q)
\]
where $|\mathcal{H}_{q'}|$ is the number of subgraphs retrieved for query $q'$, and the total number of subgraphs is the denominator $\sum_{q' \in \mathcal{Q}} |\mathcal{H}_{q'}|$. The term $\text{Metric}(H, q)$ represents the metric value for a specific subgraph $H$ that was retrieved for query $q$. The calculation for a single subgraph $H=(V_H, E_H)$ is detailed below.

\textbf{QRScore}
The QRScore quantifies the semantic alignment between a subgraph and the query that retrieved it. It's the same as Equation~\ref {eqScore}.

\textbf{Density} measures the internal structural cohesiveness of a subgraph $H$. It is the ratio of existing edges to the maximum possible number of edges for its set of nodes
    \[\text{Density}(H) = {|E_H|}/(|V_H|(|V_H| - 1)).\]
    
\textbf{Diameter} reflects the structural compactness of the knowledge within $H$. It is defined based on shortest paths within the original global graph $G$. The diameter of a retrieved subgraph $H$ is the maximum shortest path distance between any pair of its nodes that are reachable in $G$.
\[
    \text{Diameter}(H) = \max_{\substack{u, v \in V_H \\ d_G(u, v) < \infty}} d_G(u, v)
\]
where $d_G(u, v)$ is the shortest path distance between $u$ and $v$ in $G$.

\textbf{Average Pairwise Node Similarity} assesses the internal semantic coherence of a subgraph $H$ by averaging the cosine similarity over all unique node pairs within it.
\[
    \text{AvgSim}(H) = \frac{1}{\binom{|V_H|}{2}} \sum_{u, v \in V_H, u \neq v} \cos(A(u), A(v))
\]

\end{document}